\documentclass[journal=jpcbfk,manuscript=article]{achemso}
\usepackage{chemformula} 
\usepackage[T1]{fontenc} 
\usepackage{amsmath}
\usepackage{graphicx}
\usepackage{pslatex}
\usepackage{color}
\usepackage[normalem]{ulem}
\usepackage{float}
\usepackage{xcolor}  

\title{Kinetic Model of Translational Autoregulation}

\author{Vivian Tyng and Michael E. Kellman}
\email{kellman@uoregon.edu}
\affiliation{Department of Chemistry and Biochemistry and Institute of Theoretical Science, University of Oregon \\ Eugene, OR 97403, USA}
\date{\today}

\begin{document}

\begin{abstract}

We investigate dynamics of a kinetic model of inhibitory autoregulation as exemplified when a protein inhibits its own production by interfering with its messenger RNA, known in molecular biology as translational autoregulation.   We first show how linear models without feedback set the stage with a nonequilibrium steady state that constitutes the target of the regulation.  However, regulation in  the simple linear model is far from optimal.    The negative feedback mechanism whereby the protein ``jams" the mRNA greatly enhances the effectiveness of the control, with response to perturbation that is targeted, rapid, and metabolically efficient.  Understanding the full dynamics of the system phase space is essential to understanding the autoregulation process.    

\end{abstract}

\maketitle

\section{Introduction}

Autoregulation is extremely important in a multitude of contexts.   Examples range from the molecular level of gene regulation, to organ level control of physiological processes, e.g. control of  
blood flow under blood pressure variation \cite{blood}.  Perhaps the most common type of autoregulation is negative (or repressive or inhibitory) regulation in an activator-repressor network \cite{ActivatorRepressor}, represented as follows in a very simple network diagram shown in Fig. \ref{activatorrepressor}:  

\begin{figure}[hbtp]
\caption{Schematic activator-repressor network. \label{activatorrepressor}}
\begin{center}
\includegraphics[width=2.6in]{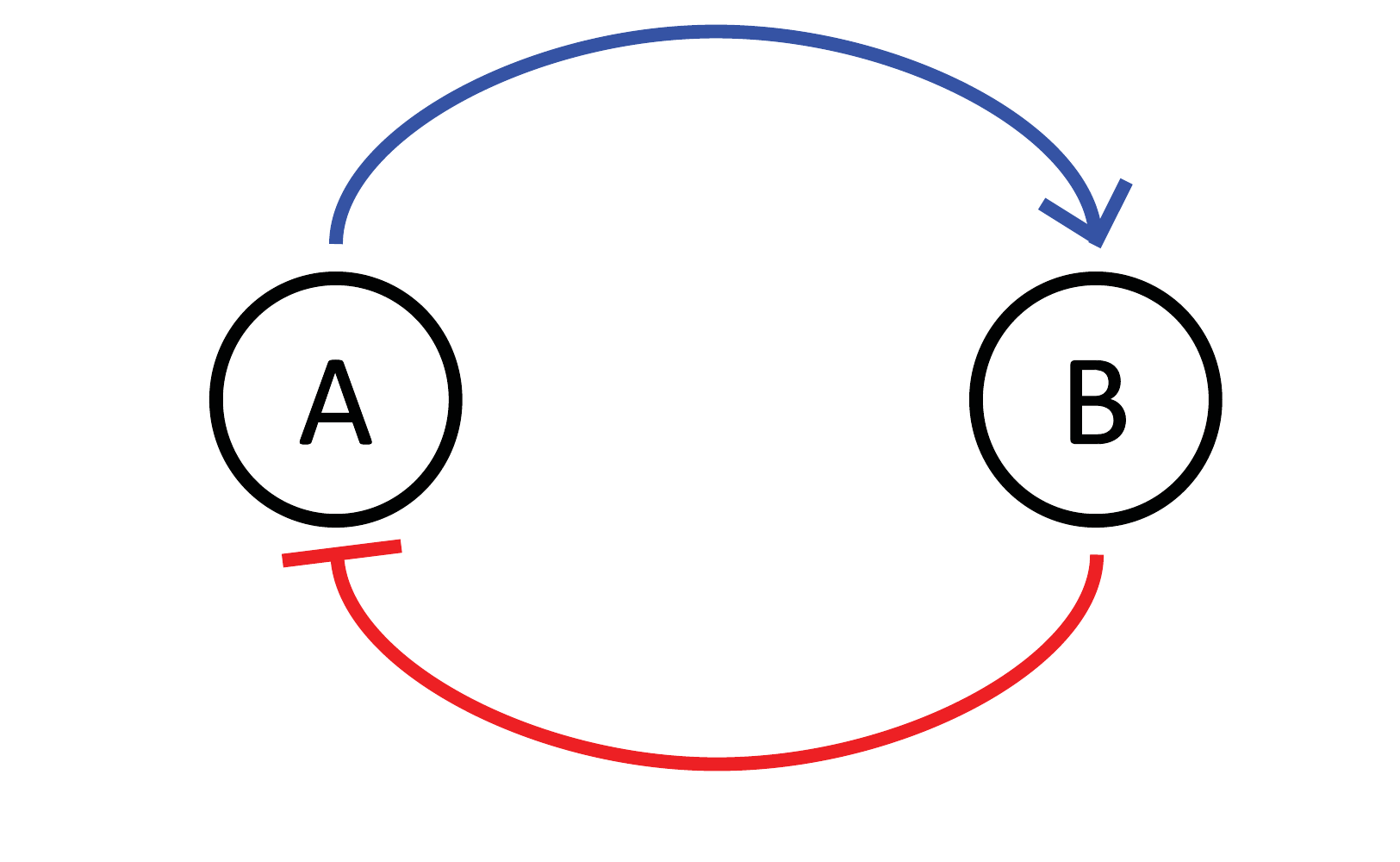}
\end{center}    
\end{figure}

\noindent  with the  arrow conventionally  representing activation and the block repression or negative feedback.  

\begin{figure}[H]
\caption{Transcriptional and translational autoregulation of gene expression.}\label{DNARNAprotein}
\begin{center}
\includegraphics[width=3.2in]{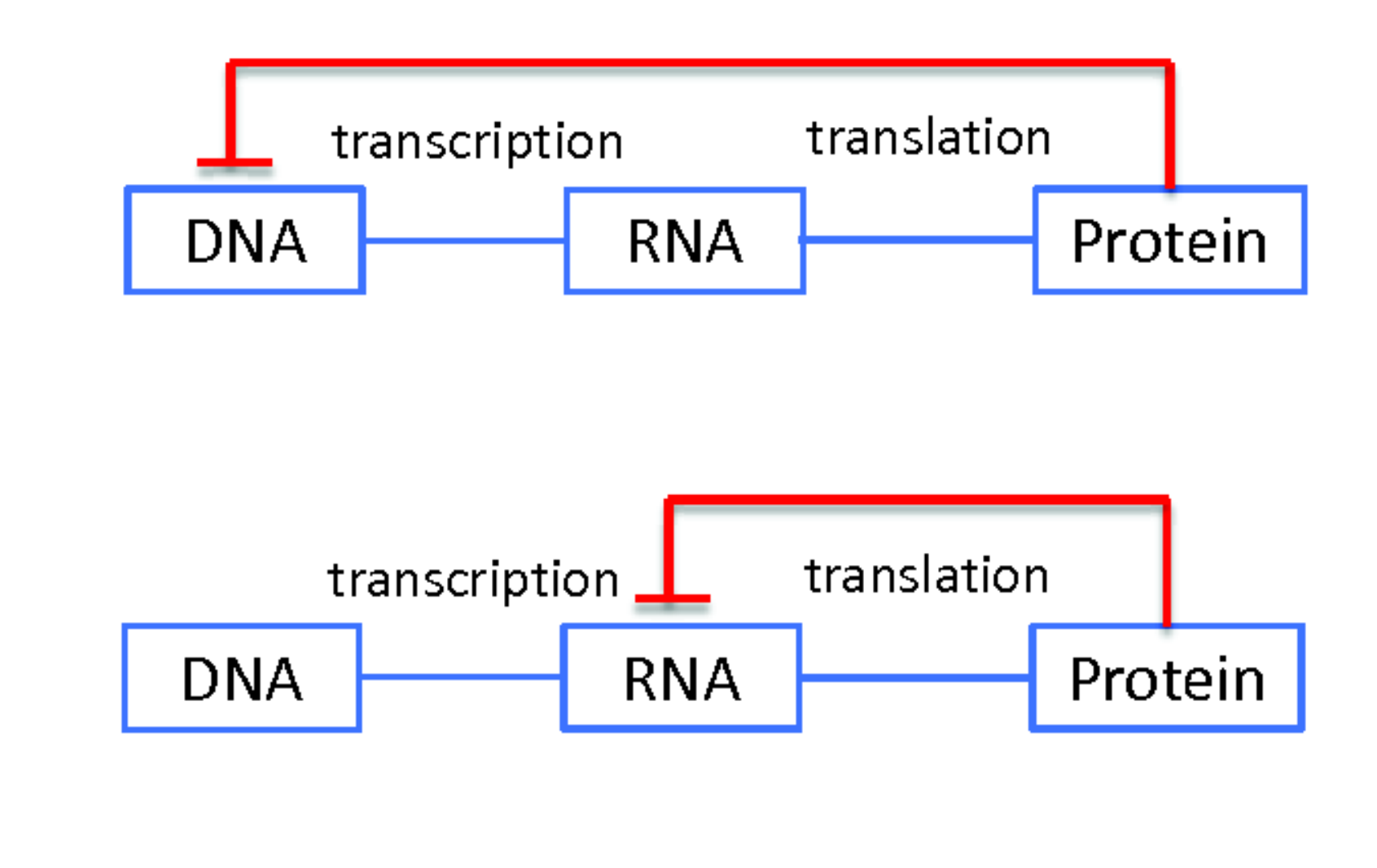}
\end{center}
\end{figure}

A very important example in  molecular biology is when a protein regulates its own production by inhibiting the translation of its gene into mRNA.  This is {\it transcriptional autoregulation}.    In another  important mechanism,  the subject of  the investigation here,  the control takes place through binding of the protein to the mRNA produced by the gene for the production of the protein.  This mechanism wherein the  protein ``jams" its own mRNA template is called {\it translational autoregulation}.   These epigenetic mechanisms of autoregulation take place among the elements of the ``central dogma" of molecular biology that ``DNA makes RNA makes Protein."  A schematic is shown in Fig. \ref{DNARNAprotein}.   Examples of translational autoregulation, which has proven to be a widespread phenomenon \cite{GlobalMammalian}, range from the gp32 protein \cite{Lemaire,vonHippel1982,Shamoo,T4Book} involved  in the  DNA replication process of T4 virus in {\it E. coli}, to thymidylate synthase (TS), a protein that plays important roles in a variety of common cancers \cite{Hallmark}.

\section{Kinetic model for translational autoregulation with negative feedback}

In this section we consider the motivation for detailed kinetic analysis of autoregulation,  then describe the elements of  the specific model of  translational autoregulation that is the focus of this paper. 

\subsection{Why a kinetic model?}

Translational autoregulation is an example of a process in a far-from equilibrium system, and as such has intrinsic interest as a problem in kinetics.    Moreover, kinetic analysis is recognized as an essential element for understanding of biologically crucial gene regulation processes \cite{Ingalls,AlonBook,Hargrove}.  Protein and RNA kinetics are becoming more accessible with advances in proteomics and transcriptomics \cite{Marcotte}.   We seek to advance understanding of the protein-mRNA regulatory and kinetic problem by adding to earlier models of this system 
\cite{Hargrove} the additional fundamental element of nonlinear feedback -- a combination which has not to our knowledge  been exploited in the mathematical analysis of translational autoregulation kinetics, though feedback is most certainly known as an essential element of mathematical systems biology \cite{Ingalls,AlonBook}.  

A kinetic model gives the {\it sequence in time} of the concentrations of all the species in  the regulatory system \cite{Houston}.  Here we will be studying the full dynamics of a two-component model.  
A simple qualitative characterization like the network diagram in Fig. \ref{activatorrepressor}  simply cannot do full justice to the regulatory process.  The purpose of our investigation of a kinetic model of  autoregulation is to understand how the detailed quantitative dynamics of the system of rate constants with negative feedback and cooperative behavior sets the stage for this particular type of autoregulation, and optimizes its behavior.  

We will confine our attention here to the translational autoregulatory process.  The DNA transcriptional mechanism, already considered by Rosenfeld et al. \cite{Rosenfeld} but in a lower-dimensional approach than we adopt, has other features that deserve a separate treatment.      Others \cite{Hargrove,Marcotte} have considered two-component models of translational regulation that do not however include feedback.   These early models are formally identical to the ``linear" models that will be our starting point here.   Hargrove and Schmidt \cite{Hargrove} argue as we do for the great importance of these simple  models for beginning to understand gene regulatory processes at the mathematical systems level.  Among our central goals are to see how the phase space view of dynamics adds greatly to understanding of even the simple linear systems.   Ref. \cite{Hargrove}  considered the individual concentration changes vs. time for specific initial conditions, but did not plot the two-dimensional dynamics in a phase space ``portrait" in the vicinity of  the steady state, i.e. the target of the negative autoregulation.  We find that this helps greatly to understand the linear systems, and then to see how the addition of feedback contributes immensely to the possibilities for optimizing the autoregulatory control.

\subsection{Elements of an effective control scheme}  

\begin{figure}[H]
\caption{Schematic of the translational autoregulatory model including feedback.  m: mRNA, P: protein.} \label{autoregulationscheme}
\begin{center}
\includegraphics[width=3.5in]{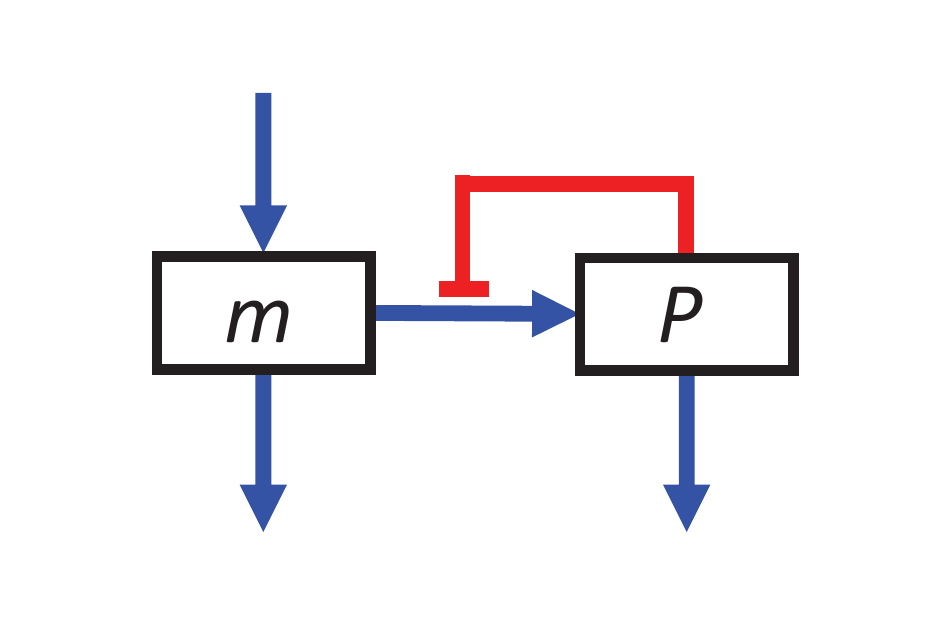}
\end{center}
\end{figure}

To build an effective control scheme, we need a system in which the control is {\it targeted} and {\it rapid}. 
We may also want to take into account the {\it metabolic energy demands} of various possible kinetic schemes (including their parameters) on an organism, aiming for control that is {\it efficient}.     
To achieve all of this, we need two elements, as shown in Fig. \ref{autoregulationscheme}: (1) a basic ``barebones"  linear kinetic scheme of reactions without feedback.  This has rates for production (via gene transcription) and degradation of the mRNA; and production (via translation of the mRNA by the ribosome machinery) and degradation of the protein. As emphasized  quite some time ago by Hargrove and Schmidt \cite{Hargrove}, this is already a simple regulatory system that has the feature of a unique steady state that is the basic target of the regulation, which dampens deviations from the steady state.  
 (2)  Then,  we need additional features of  feedback and cooperativity (terms defined mathematically below) that give to the autoregulation the robust aspects of control outlined above  (``targeted, rapid, efficient").    The feedback and cooperativity ``tune" the  barebones linear network to give superior performance.     Fig. \ref{autoregulationscheme}  shows the autoregulatory network with the species  mRNA ($m$) and protein ($P$), with arrows representing the barebones linear scheme, and the negative feedback with cooperativity depicted as the blocking by $P$ of its own production with $m$.    

In the following sections, we will build the barebones linear model and examine its essential features and  limitations of performance, then tune this model with feedback and cooperativity to see how this gives superior  regulation.  We will see that each component of the system affects the performance and the tuning of the other components.  Our plan is the following.   First, in Section \ref{linear} we imagine a cell or virus that seeks a certain steady state level of the protein $P$.  We construct  a linear model that gives the desired concentration.  We will see that there is a natural classification of the dynamics into three qualitatively different types in the linear model, depending on a key ratio of two of the parameters.    Then, in Section \ref{feedback} we imagine the cell or organism adapting by adding feedback and cooperativity to get better autoregulatory control (e.g. faster response times) than is afforded by the linear model, while maintaining the same target steady state level of $P$.  We will choose a numerical ``feedback strength,"  and adjust the other parameters accordingly.      We will compare the behavior  of each of the three types of control dynamics with feedback with that of the corresponding linear system.   Then, in Section \ref{systematic} we make a systematic comparison of the autoregulatory control performance of the linear systems and the corresponding systems with feedback. 

\section {Linear model with steady state}    \label{linear}    

We imagine a cell or virus that seeks a certain steady state level $P_{SS}$ of the protein, and systematically construct  a linear model, without feedback, that would accomplish this goal.  

This ``barebones" linear model already represents the simplest case of a control scheme.  Great insight can be obtained because of its simplicity.  It needs to be stated that the linear model has been considered long ago  and its essential importance in protein regulation recognized;   our equations    (\ref{lineareqns}-\ref{sssolution})  are identical in content to the system of Ref. \cite{Hargrove}. However, our phase space portraits with organization into three classes are new, as is our addition later of nonlinear feedback as an essential element.   Another difference is that Ref. \cite{Hargrove} emphasizes analytical solutions for the linear model, while we also emphasize numerically computed dynamics for both the linear system and the system with nonlinear feedback, as is essential for the latter.  As noted already, we find that there are three qualitatively different types of linear models, depending on the chosen parameters.   {We will examine the control properties of each of these versions of the model, with a view toward their individual advantages and shortcomings.  }

\begin{figure}[H]
\caption{Linear model without feedback} \label{linearmodel}
\begin{center}
\includegraphics[width=3.5in]{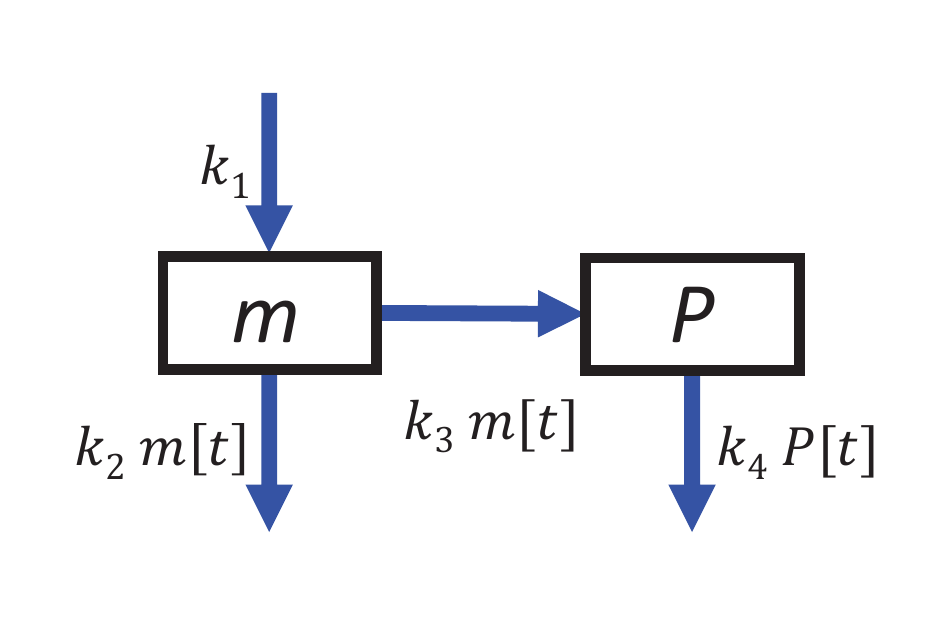}
\end{center}
\end{figure}

The model  has two  species $m$ and $P$ and four  rate constants $k_1 ...  k_4$, as shown in Fig \ref{linearmodel}.  $k_1$ gives the rate of production of $m$ via transcription of the gene for the protein $P$; $k_2$ is a rate constant for degradation of $m$;   $k_3$ is the rate constant for translation of the mRNA into  $P$; $k_4$ is the rate constant for  degradation of $P$.  As we shall see, Hargrove and Schmidt did not plot the two-dimensional dynamics in a phase space ``portrait". The relative degradation rates of $m$ and $P$ turn out, perhaps surprisingly, to be a  key  determinant of the type of dynamics that is obtained.  The kinetic equations  are  

\begin{eqnarray}
\frac{dm}{dt} = k_1-k_2 m, \quad \quad \frac{dP}{dt} = k_3 m-k_4 P \label{lineareqns}
\end{eqnarray}

\noindent There is a single steady state (SS) solution when the net change rates of both species are zero: 

\begin{eqnarray}
m_{SS}=\frac{k_1}{k_2}, \quad \quad P_{SS}=\frac{k_1 k_3}{k_2 k_4} = m_{SS}\frac{k_3}{k_4}   \label{sssolution}
\end{eqnarray}

\noindent    This single steady state is the basic target of the control system -- we will verify later that the addition of feedback, though nonlinear, does not result in bifurcations to more steady states.  Our model begins by specifying $m_{SS}, P_{SS}$ i.e. the parameter ratios in Eq. \ref{sssolution}.  

Next, we compute the dynamics around the steady state on a diagram of {\it normalized} concentrations $m/m_{SS}, P/P_{SS}$, with the steady state concentrations at (1, 1).   We choose values of the rate constants at will to give a particular instantiation of the model.   As the first example,  we pick $k_1 = 20/17, \ k_2 = 10, \ k_3 = 34, \ k_4 = 1$.  These parameters can be found as the first set in Table \ref{TableI}. 
The flow in the    phase space of ($m/m_{SS}, P/P_{SS}$) is  shown in Fig. \ref{3linearpatterns}(a).     We see something very interesting.  There is  a ``vertical structure" in the flow toward the steady state.  
There is first a general nearly horizontal fast flow toward the vertical.   Then,   flow takes place much more slowly asymptotically to the vertical toward the steady state.  This latter flow is said to be along the  vertical ``slow manifold" \cite{MaasPope,DavisSkodje}. This can be understood in terms of the  ``linearized flow" near the SS in standard nonlinear dynamical analysis using the  Jacobian; we go into detail in the Appendix.  Dynamically, this behavior stems from the fact that the mRNA turnover rate is much faster than the protein turnover rate ($k_2/k_4 \gg 1$). As a result, along the fast manifold, the mRNA concentration quickly reaches the steady state value $k_1/k_2$ and remains so, followed by the relatively slower change in $P$. This kind of dynamics can be described by the ``quasi-steady state approximation" in traditional chemical kinetics \cite{Houston}.

Investigating various parameter sets, we find that there are three general patterns of flow, which we designate in Fig. \ref{3linearpatterns} as ``vertical," ``focus," and ``diagonal."    The {corresponding} parameters  are listed in the upper left part designated ``linear" of Table \ref{TableI}.    All of the flows have a  separatrix along the vertical at {$m = m_{SS}$} , resulting from a vertical eigenvector associated with the Jacobian matrix, as discussed in the Appendix.  (In the vertical case, Fig. \ref{3linearpatterns}a, the separatrix corresponds to the slow manifold.)  The patterns correspond to  the following relations among the kinetic parameters:

\

1)  Vertical case in Fig. \ref{3linearpatterns}a, $k_2 \gg k_4, (k_2/k_4 = 10$ from Table \ref{TableI}):  {Fast relaxation of $m$ (horizontal trajectories) to the steady state value, followed by slow change in protein concentration (vertical slow manifold). As mentioned earlier, this corresponds to the ``quasi-steady state approximation"  of traditional kinetics. Rosenfeld et al. \cite{Rosenfeld} made use of this approximation for the translational regulation case.

2)  Focus case in Fig. \ref{3linearpatterns}b, $k_2  \approx   k_4$ ($k_2/k_4 = 1$ from Table \ref{TableI}) The eigenvalues $-k_2$, $-k_4$ are comparable, the trajectories are spiral-like  so there is no separation in timescale into fast and slow manifolds. A similar case had been considered by Nov\'{a}k and Tyson \cite{Novak}.       

3)  Diagonal case in Fig. \ref{3linearpatterns}c, $k_2 \ll k_4$,  ($k_2/k_4 = 0.1$ from Table \ref{TableI}):  Here, the mRNA turnover rate $k_2$ is much {\it slower} than the protein turnover rate $k_4$. 

\

It is very interesting that these basic patterns of the normalized phase portraits are determined by the ratio $k_2/k_4$ of {\it degradation rates} of the mRNA and the protein.  The importance of the degradation rates was emphasized by Hargrove and Schmidt \cite{Hargrove}.  (We will see in what follows that these three basic patterns play a crucial role in the classification of dynamics that the control system can exhibit.)  Further, it turns out that the normalized ($m_{SS}$,  $P_{SS}$) phase flows {are a universal property of the linear model that} depend only on the  ratio $k_2/k_4$, invariant under any change in the other three variables in the parameter space.   To understand this  property mathematically, note that the linear partial differential equations in (\ref{lineareqns}) can be solved analytically:

\begin{eqnarray}
m(t) &&= \frac{k_1}{k_2}+C_1 e^{-k_2 t} \nonumber\\
P(t) && = \frac{k_1 k_3}{k_2 k_4}+\frac{C_1 k_3 (e^{-k_4 t}-e^{-k_2
t})}{k_2-k_4}+C_2 e^{-k_4 t}        \label{lineartimedep}
\end{eqnarray}
These depend only on $k_2, k_4$. 
This has the very important consequence that {\it the normalized phase flow depends only on the ratio $k_2/k_4$.    }
When, as we will consider later,  the initial condition satisfies $m[0]=
m_{SS}$, we have $C_1=0, m(t) = m_{SS}$. Then the time-dependent protein
concentration can be simplified as:

\begin{eqnarray}
P(t) = \frac{k_1 k_3}{k_2 k_4}+C_2 e^{-k_4 t}.       \label{Pvertical}
\end{eqnarray}

\noindent Clearly, in this special case,   $P(t)$  temporally depends only on $k_4$. 

\begin{figure}[H]
\caption{Representative normalized phase portraits
for (a) vertical, (b) focus, and (c) diagonal types. {The time step (space between the dots) is the same for all panels.}} \label{3linearpatterns}
\begin{center}
\includegraphics[width=2.1in]{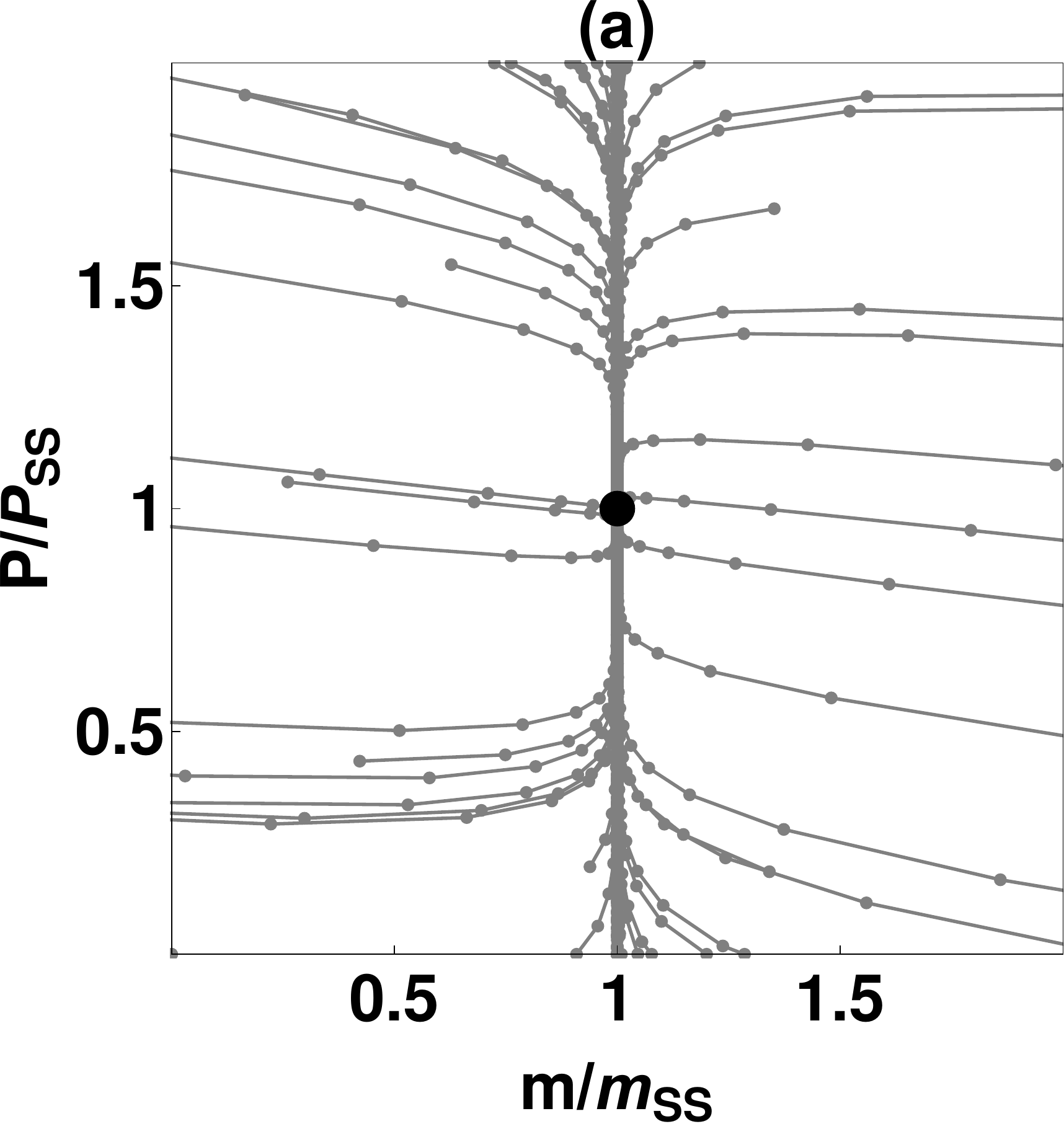}
\includegraphics[width=2.1in]{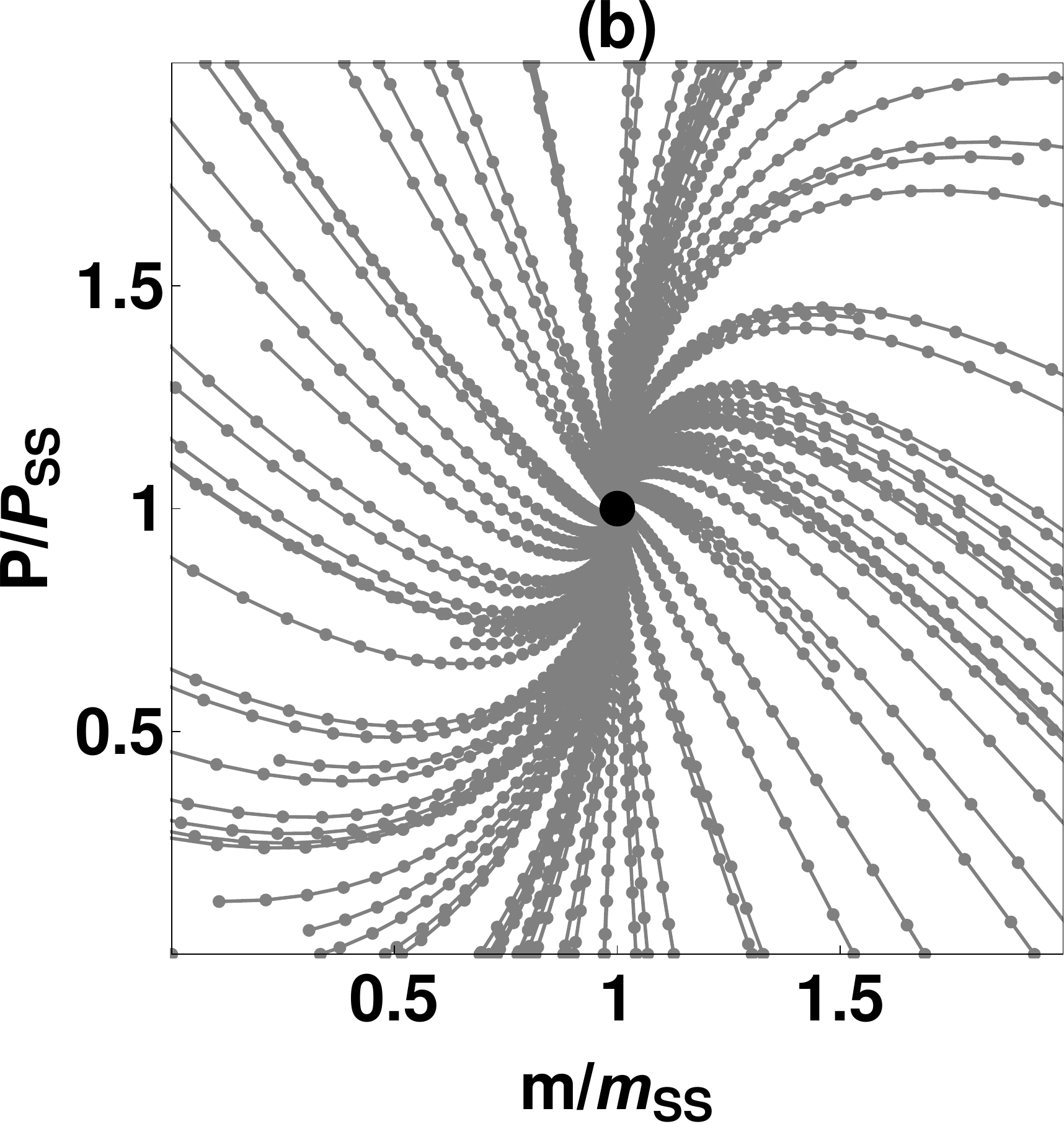}
\includegraphics[width=2.1in]{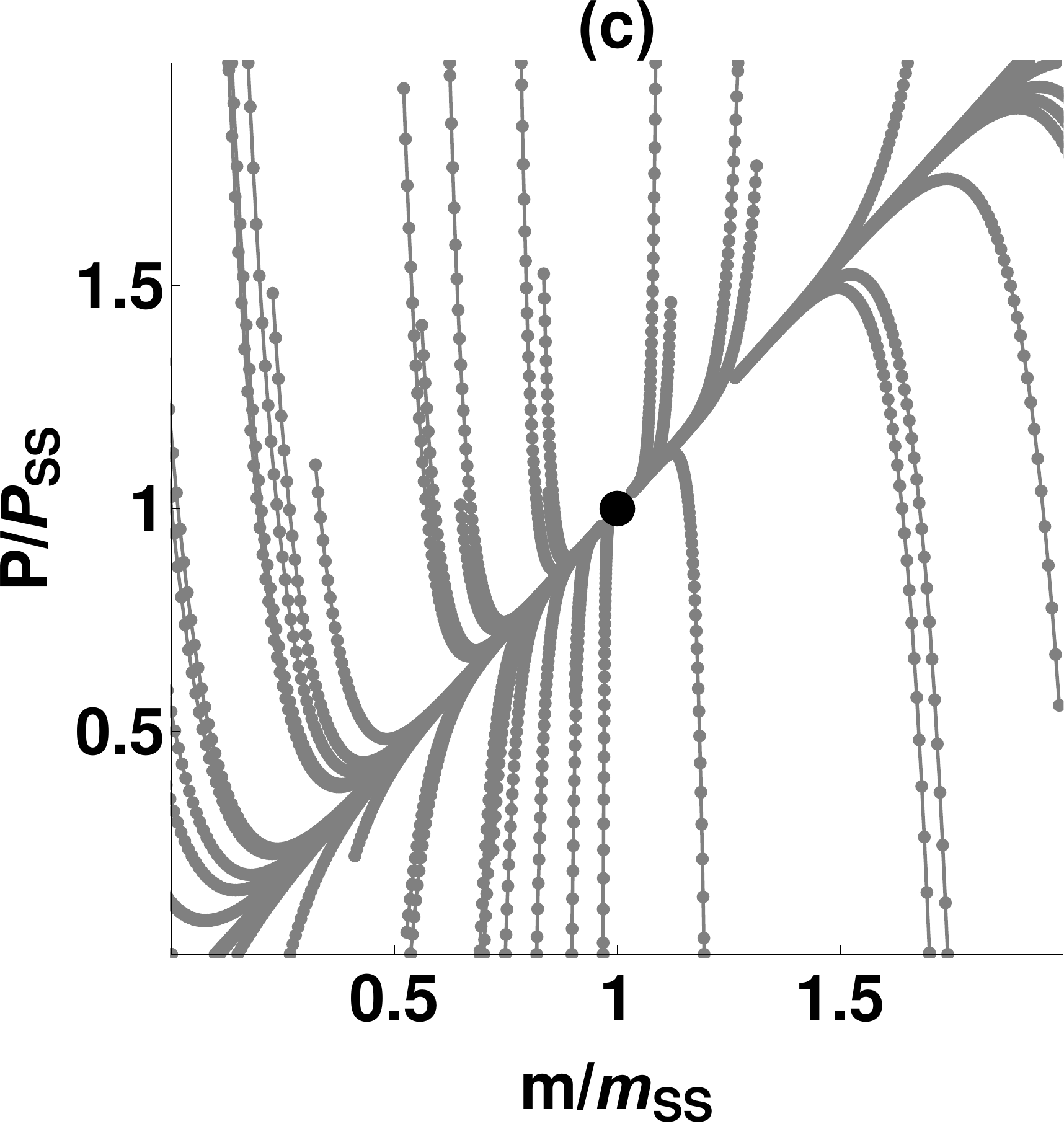}
\end{center}    
\end{figure}

\section {Adding Feedback and cooperativity to tune the  autoregulatory system.}  \label{feedback}

Already, these linear models constitute  basic control schemes: in each, there is a steady state that is an attractor and the desired target for the dynamical system.  However, this is a very primitive kind of control.  Only the vertical profile has fast control directly to the $P_{SS}$.  The focus profile has wandering  trajectories with arcing excursions.  Even worse, the diagonal profile quickly directs the trajectory to the diagonal manifold, which however does not in general have $P$ near the steady state value, so the trajectory gets stuck away from the target for a relatively long time.  Moreover, the  possibility, such as it is in each case, of achieving quick protein control  by speeding up all the rates, also is extravagantly wasteful.   After building $m$ and $P$ at great energetic cost, they are degraded, again at great cost!  It is like filling buckets -- one for $m$, a second, hydraulically linked one for $P$ --  that have holes in them designed to regulate the level in each bucket.  The holes are the degradation processes with rate constants $k_2 $ and $k_4$.    To speed the system by a factor $\alpha$, it suffices to increase all the parameters by $\alpha$.  This is certainly effective, but the metabolic cost  just as certainly is extravagant. To switch metaphors, the processes of production of $m$ and $P$ are like an accelerator on a car, and the processes of degradation are like brakes.  It is desirable to have a more subtle and discriminating system of  accelerator and brakes.  This is achieved by adding the element of nonlinear feedback with cooperativity.  We will demonstrate these statements in the rest of the paper.  

\begin{figure}[H]
\caption{Schematic of autoregulation model with feedback and cooperativity, with rate constants and Hill-type factor.} \label{feedbackmodel}
\begin{center}
\includegraphics[width=3.5in]{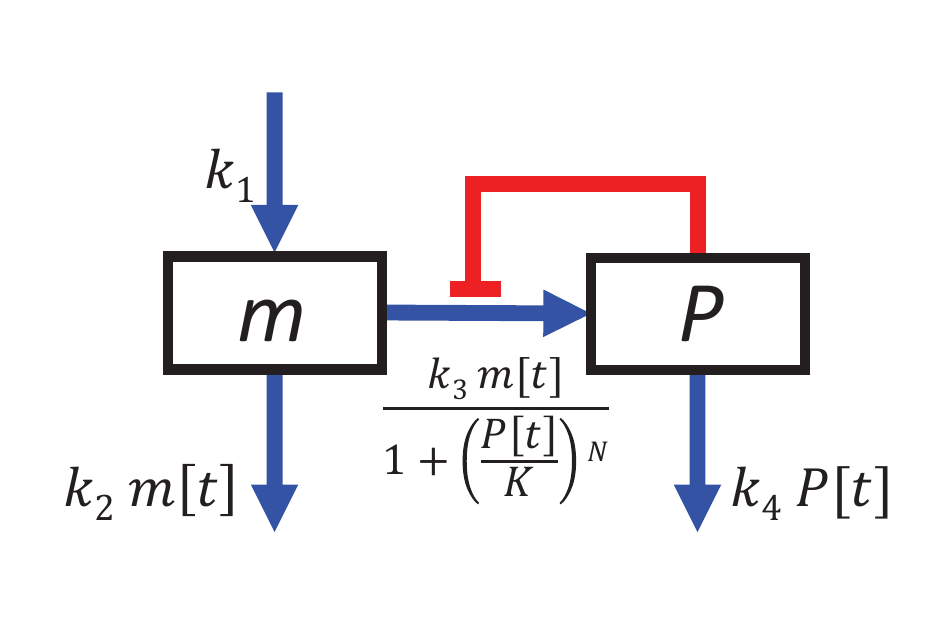}
\end{center}
\end{figure}

We adopt a model with typical features of feedback and control, shown in Fig. \ref{feedbackmodel}, modifying the protein synthesis rate using  ``Hill-type" parameters often associated \cite{Ingalls} with feedback $K$ and cooperativity $N$:

\begin{eqnarray}
\frac{dm}{dt} = k_1-k_2 m = 0, \quad \quad \frac{dP}{dt} = \frac{k_3 m}{1+(P/K)^N}-k_4 P = 0        \label{feedbackmodeleqn}
\end{eqnarray}

\noindent  {The denominator in the $P$ rate equation is the feedback factor $X$, a function of $P$ given by}

\begin{equation}
 X(P) = 1+(P/K)^N         \label{hilltypeX}
\end{equation}

\noindent and  is intended to represent the jamming of the mRNA by binding to its product protein $P$.  The factor $X(P)$ is a typical Hill-type expression involving the feedback parameter $K$ and the cooperativity parameter $N$.  We regard this kind of term in the way it is often used, as a phenomenological or empirical expression, not a literal expression in terms of a binding parameter $K$ and cooperativity number $N$. 
 The origin in enzyme kinetics and expanded empirical use of the Hill-type expressions is discussed in great detail by Ingalls \cite{Ingalls}.  

In effect, the concentration of available $m$ is reduced by the factor $1/X(P)$. 
 To represent this in the model, first we pick a value $X = X(P_{SS})$   at the steady state.   This modifies the expression for $P_{SS}$ from Eq. \ref{sssolution}  to  

\begin{eqnarray}
m_{SS}=\frac{k_1}{k_2}, \quad \quad P_{SS}=\frac{k_1 k_3}{k_2 k_4}\frac{1}{X} = m_{SS}\frac{k_3}{k_4}\frac{1}{X}    \label{sssolutionfb}
\end{eqnarray}

\noindent {To maintain the same value of $P_{SS}$ at the steady state, which we take to be the ``goal" of the organism in ``designing"  the control scheme, with or without feedback, we need to compensate for $1/X(P)$ in equations (\ref{feedbackmodeleqn},\ref{hilltypeX}).  To do this, we choose to enhance the gene transcription rate constant  $k_1$ by the factor $X$ -- this seems the most likely of many possible scenarios involving the $k_i$.  This increases $m_{SS}$ from its value in the linear model by the factor $X$, to }

\begin{equation}
m'_{SS} = \frac{k_1}{k_2}X    \label {mssprime}
\end{equation}

\noindent     In effect, to maintain $P_{SS}$ with the feedback, the system maintains a ``reserve army" of messenger RNA via the enhanced rate constant $k_1$.

\section{Systematic analysis of the autoregulatory system}      \label{systematic}

So far we have established the basic framework of the linear system, with classification of the dynamics into three types based on the key ratio $k_2/k_4$;  and then built in feedback with cooperativity.    Now we  systematically analyze the behavior and autoregulatory performance of the system, showing that with feedback there generally is far better control than with the corresponding linear system, with better time response to perturbation of the steady state {while limiting the metabolic cost}.  We will do this in three steps:  (A) Introduce  prototypical individual trajectories from the phase space portraits, selected as being especially important examples of the autoregulatory process.  (B)  Then, consider systematic variation of the parameters in the kinetic model, in particular,  of the key ratio $k_2/k_4$   with $k_4$ fixed; and variation of this ratio with $k_2$ fixed.    (C) Examine the crucial indicator of autoregulatory efficacy, the time response of the protein concentration against various perturbations from the steady state concentrations $m_{SS}, P_{SS}$ -- first for the linear systems, and then the corresponding systems with feedback.  We will see that the feedback mechanism gives greatly improved autoregulation.   We present the tabular and visual content in Tables \ref{TableI},\ref{TableII}  and in Figs. \ref{tableIfigures},\ref{tableIIfigures},  to which we refer repeatedly in the following.    

\subsection{Prototype Trajectories}   \label{prototypes}

We will pay particular attention to  trajectories of likely special importance.      These are color coded in the figures.  One type of trajectory is where the protein concentration is perturbed from its SS value, while the mRNA concentration is unchanged.  An important example might be when the concentration of a particular protein is deliberately reduced in cancer chemotherapy \cite{Peters,TS2015Review}.  
 The purple and green trajectories in the figures have the protein concentration $P$  displaced below and above the SS value, while $m$ is kept at its steady state value, i.e. these trajectories start at  $(m, P) = (1, 2)$ and $(1, 0)$ and end at the steady state $(1, 1)$.   Another important type of trajectory likely is where both $m$ and $P$ start at zero concentration, i.e. the system is ``turning on."   This is seen in the  orange trajectories, which start at  $(m, P) = (0, 0)$.   Finally, the cyan and blue trajectories correspond to perturbation in mRNA concentration, which could occur either naturally or due to the introduction of mRNA-binding species \cite{Garg}.  

\subsection{Systematics: Variation of Degradation Rates of mRNA and Protein}

In Tables I, II and Figs. \ref{tableIfigures},\ref{tableIIfigures} we examine the effects of systematic variation in the kinetic parameters.     We want to see how the dynamics change as the system changes -- i.e. as the instantiation of our model defined by its particular parameters varies.      We have supposed that in adding feedback to a linear model, e.g. through evolution, a primary criterion for an organism might be to preserve the SS protein concentration.   Hence, we keep the steady state concentration $P_{SS} = 4$ in all cases.  We consider systematic variation of the parameters $k_2, k_4$ for degradation of $m$ and $P$ since these seem to be the key to the pattern of the dynamics.  In particular, we vary the key ratio $k_2/k_4$, first with $k_4$ fixed; then with $k_2$ fixed.   We consider linear models, without feedback; and then corresponding models with feedback added.  We will find that these  parameter variations suffice to tell us most of what we want to know about the systematic behavior of the control systems.  

In the linear models in  Fig. \ref{tableIfigures}a-c, we vary $k_2/k_4$ by keeping  $k_4$ fixed while varying $k_2$ along with $k_1$, and keeping $k_3$ fixed.   This preserves the value of the steady state concentrations $m_{SS}, P_{SS}$, according to Eqs. (\ref{lineareqns}-\ref{sssolution}).  See the corresponding values in the top part of Table I.  In the linear models in Fig. \ref{tableIIfigures}a-c, we vary $k_2/k_4$ by keeping  $k_2$ fixed along with $k_1$, while varying $k_4$ along with $k_3$.   This again preserves  $m_{SS}, P_{SS}$.  The corresponding values are given in Table II.

In the feedback models in Fig. \ref{tableIfigures}d-f, we use a feedback strength of $X_{SS} = 17$.  To compensate for this, according to Section \ref{feedback}, we increase $k_1$ from the corresponding linear models by a factor of 17.  As in the case of linear models in Fig. \ref{tableIfigures}a-c, we again vary  $k_2/k_4$ by keeping  $k_4$ fixed, while varying $k_2$ along with $k_1$.   This again preserves the value of the steady state concentrations $P_{SS}$ (while changing the value of $m_{SS}$ in  the feedback models). {The same parameter adjustment was implemented for Fig. \ref{tableIIfigures}d-f, except that now $k_2$ and $k_1$ are fixed, while $k_4$ along with $k_3$ are varied.}

These systematic variations become clearer with perusal of the tables.  

\begin{table}[H]
\caption{Parameters used for Fig. \ref{tableIfigures}. In
all cases, $k_3$ and $k_4$ are fixed; and $k_1$ varies between linear and feedback systems by the factor $X_{SS}$.}
\begin{center}
  \begin{tabular}{||c||c|c|c||}
    \hline
   \, & vertical, linear & focus, linear & diagonal, linear \\      \hline
   $k_1$ & $\frac{20}{17}$ & $\frac{2}{17}$ & $\frac{2}{170}$ \\
   $k_2$ & 10 & 1 & $\frac{1}{10}$ \\
   $k_3$ & 34 &  34 & 34 \\
   $k_4$ & 1 & 1 & 1  \\
   $K$ & - & - & - \\
   $N$ & - & - & - \\
   $X_{SS}$ & 1 & 1 & 1  \\ \hline
   $m_{SS}$ & $\frac{2}{17}$ & $\frac{2}{17}$ &
$\frac{2}{17}$  \\
   $P_{SS}$ & 4 & 4 & 4 \\
   $k_2/k_4$ & 10 & 1 & $\frac{1}{10}$  \\
$\tau_1$,\space $\tau_2$, \space $\tau_3$ & 0.693, 0.693, 0.798 & 0.693, 0.693,1.68 & 0.693,0.693,7.98 \\ \hline \hline
   \, & vertical, feedback & focus, feedback & diagonal, feedback \\      \hline
   $k_1$ & 20 & 2 & $\frac{2}{10}$ \\
   $k_2$ & 10 & 1 & $\frac{1}{10}$ \\
   $k_3$ & 34 & 34 & 34 \\
   $k_4$ & 1 & 1 & 1 \\
   $K$ & 4 & 4 & 4 \\
   $N$ & 2 & 2 & 2 \\
   $X_{SS}$ & 17 & 17 & 17 \\ \hline
   $m_{SS}$ & 2 & 2 & 2 \\
   $P_{SS}$ & 4 & 4 & 4 \\
   $k_2/k_4$ & 10 & 1 & $\frac{1}{10}$ \\
   $\tau_1$,\space $\tau_2$, \space $\tau_3$ & 0.311, 0.0361, 0.100 & 0.311, 0.0361, 0.297 & 0.311, 0.0361, 1.06 \\ \hline \hline
   $\tau_{linear}/\tau_{feedback}$ & 2.23, 19.18, 7.97 & 2.23, 19.18, 5.65 &
2.23, 19.18, 7.56 \\ \hline \hline

  \end{tabular}
\end{center}      \label{TableI}
\end{table}

\newpage

\begin{figure}[H]
\caption{{Normalized phase portraits and timecourses for $P/P_{SS}$  for Table I.}\label{tableIfigures}}
\begin{center}
\includegraphics[width=2in]{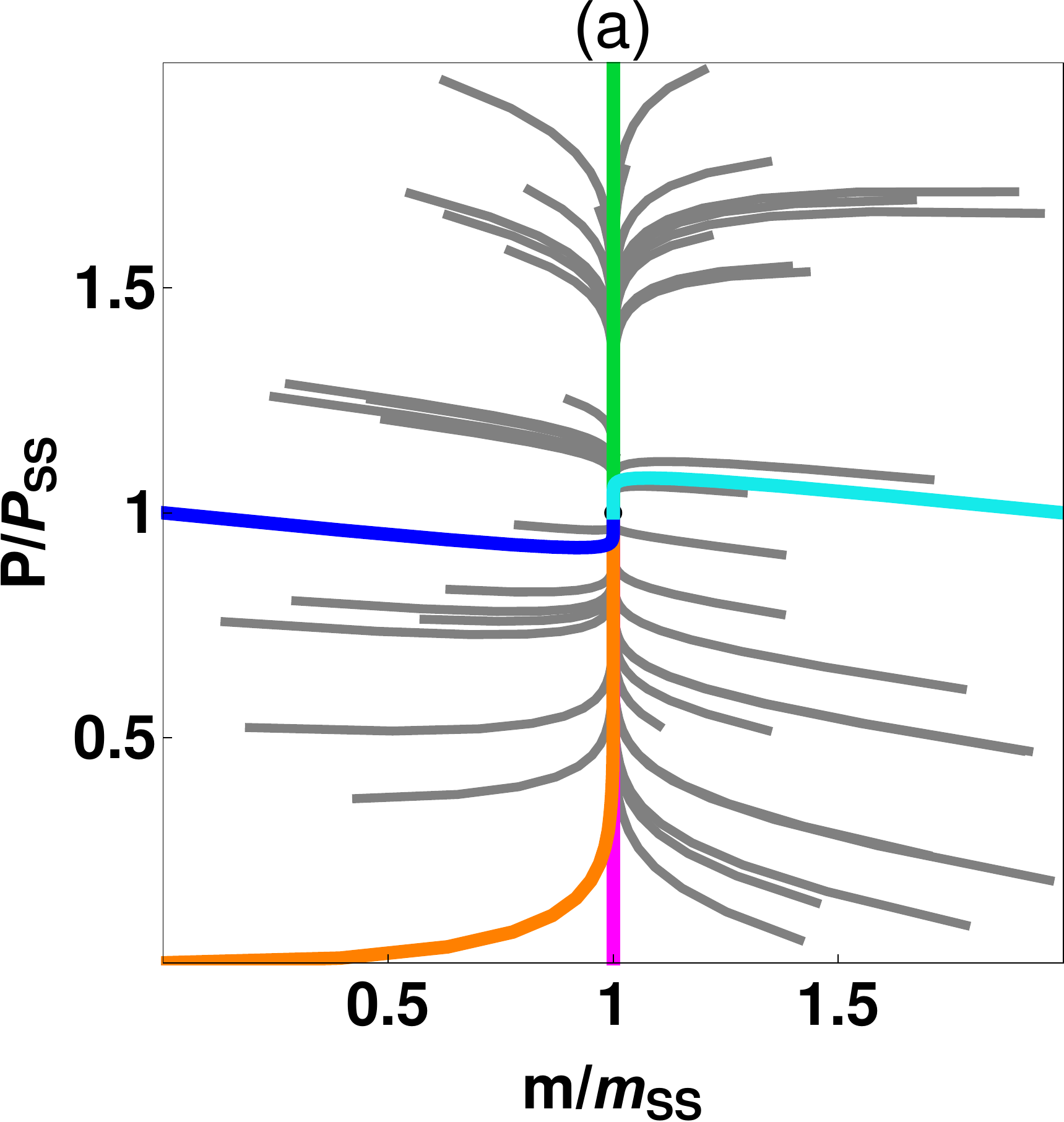}\quad
\includegraphics[width=2in]{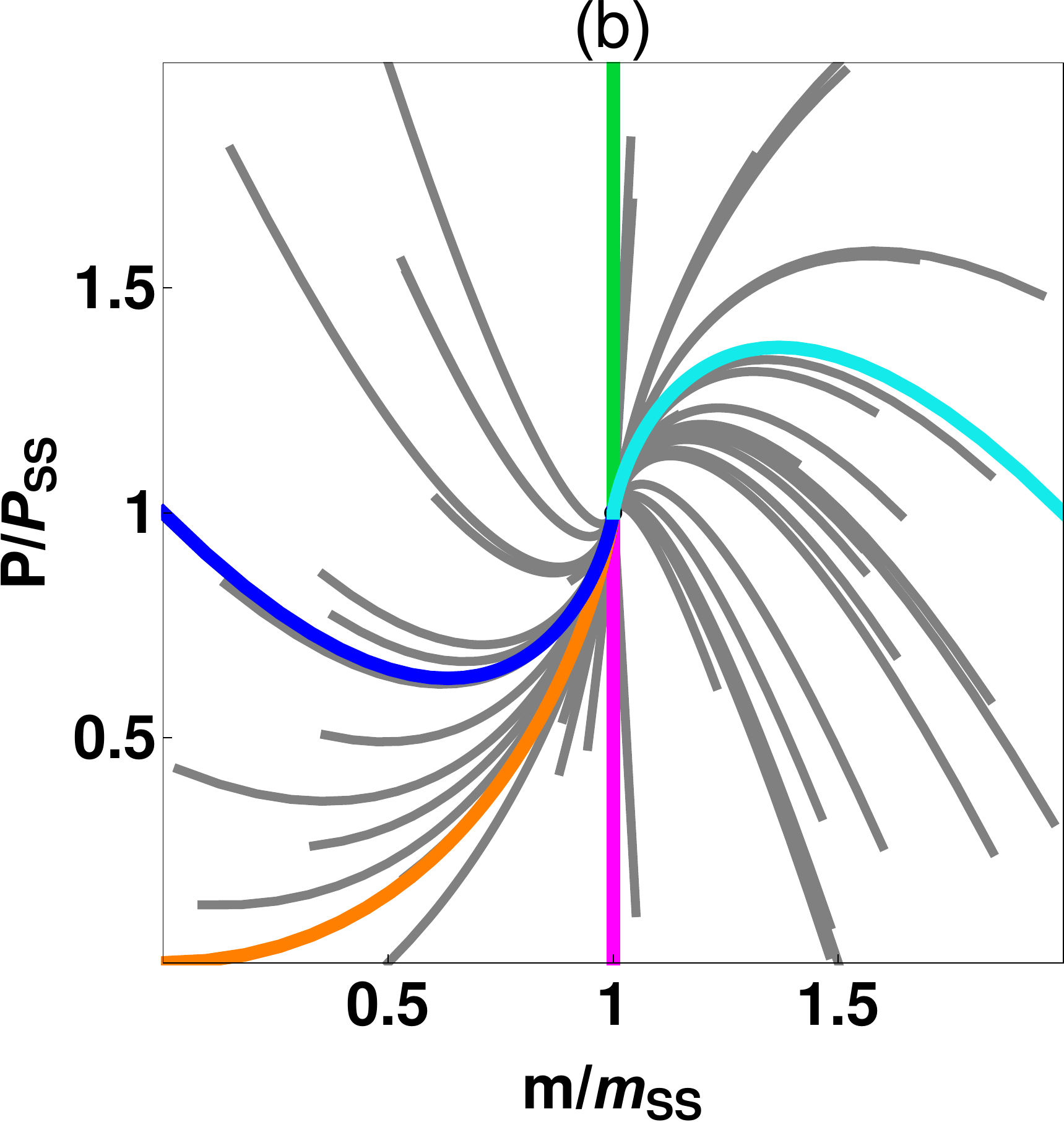}\quad
\includegraphics[width=2in]{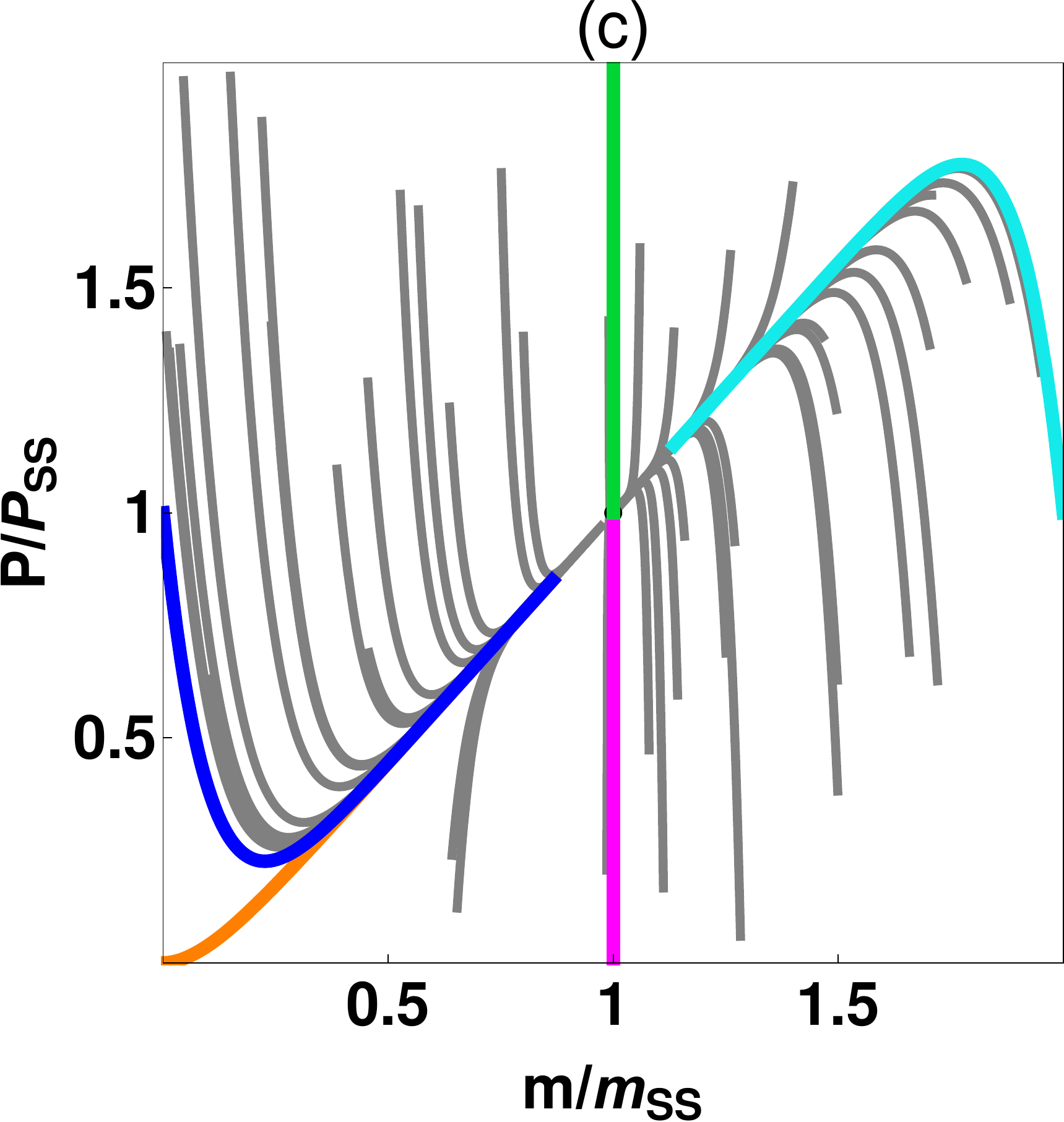} 
\medskip
\includegraphics[width=2in]{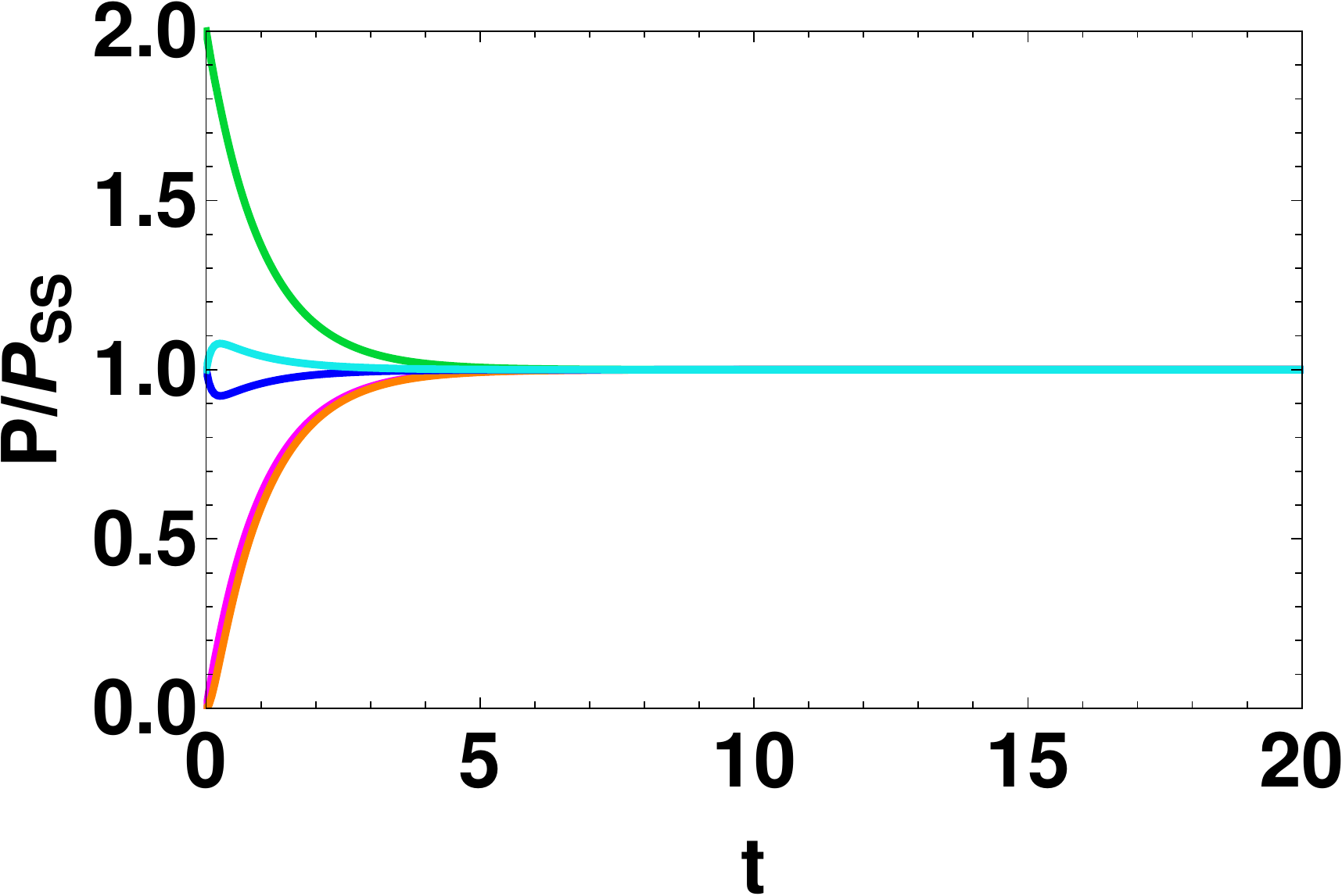}\quad
\includegraphics[width=2in]{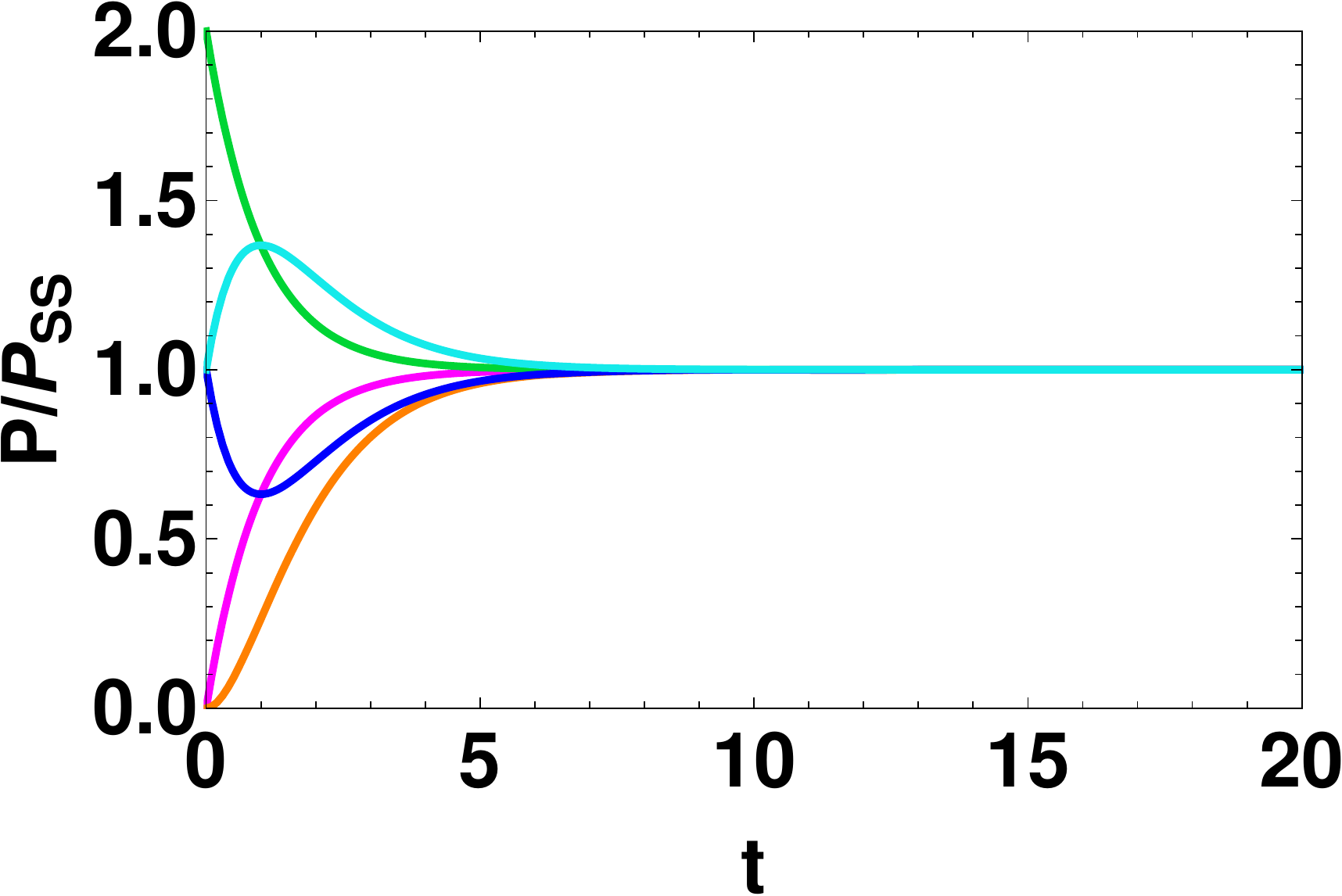}\quad
\includegraphics[width=2in]{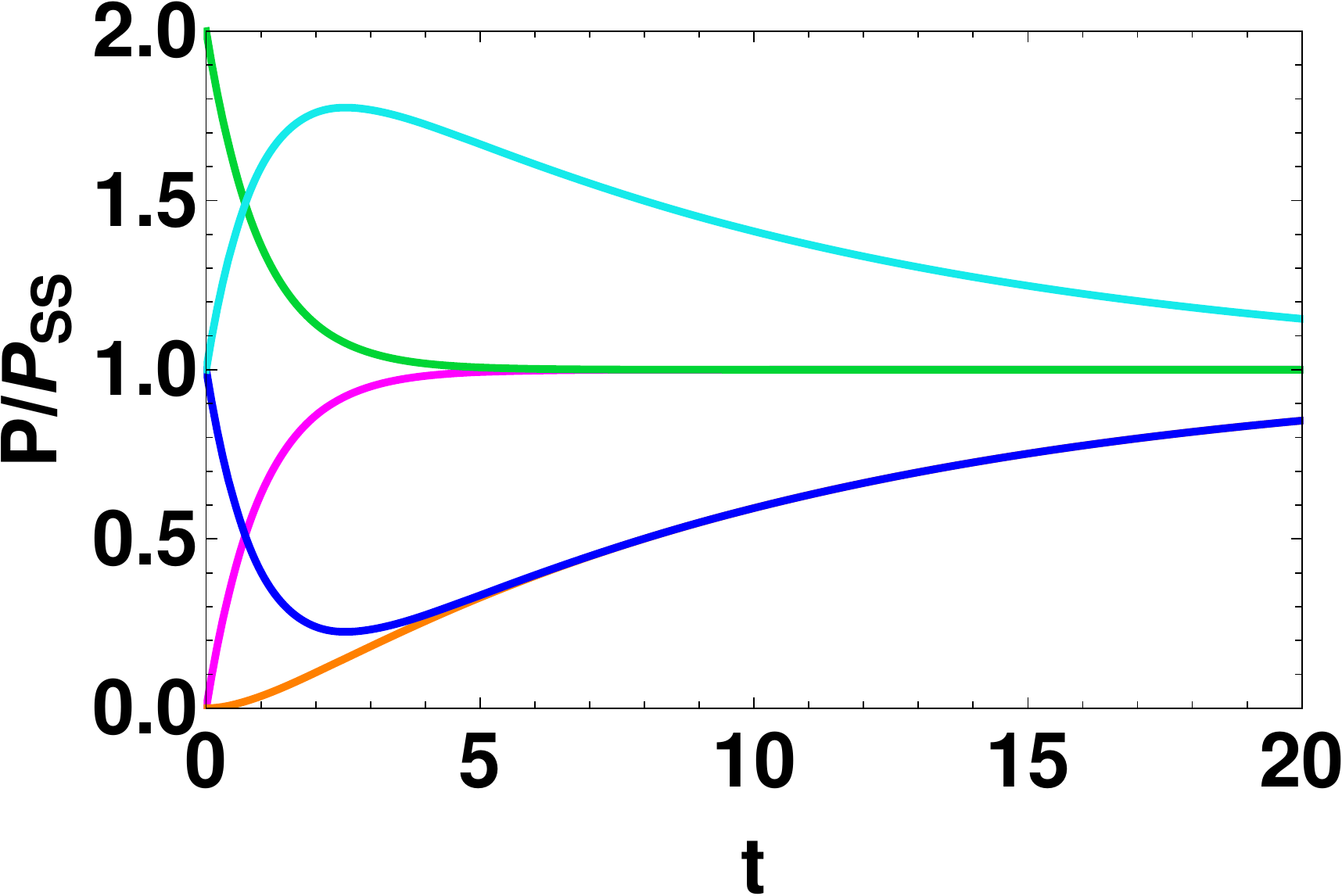} 
\medskip
\vspace{0.5in}
\includegraphics[width=2in]{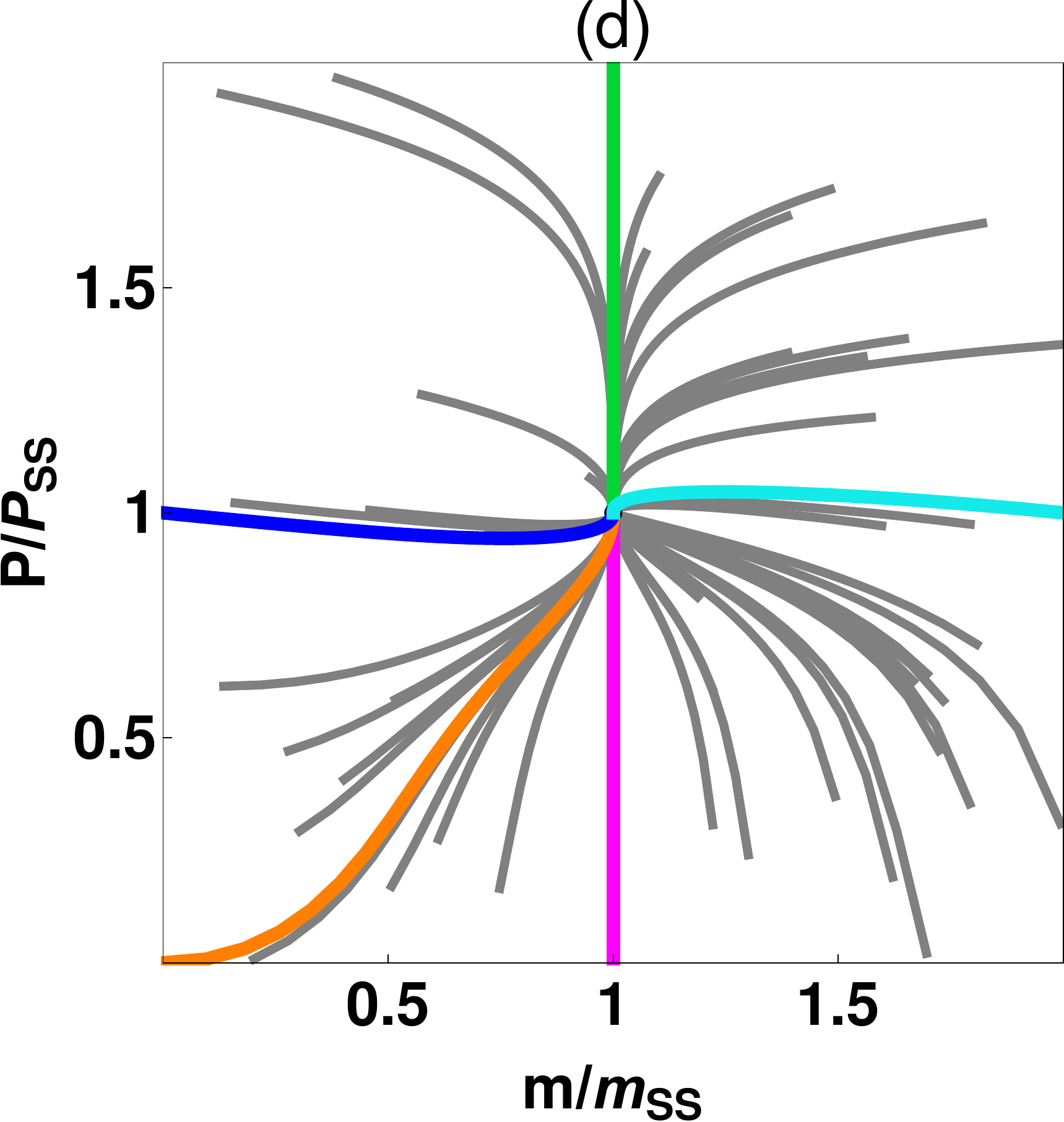}\quad
\includegraphics[width=2in]{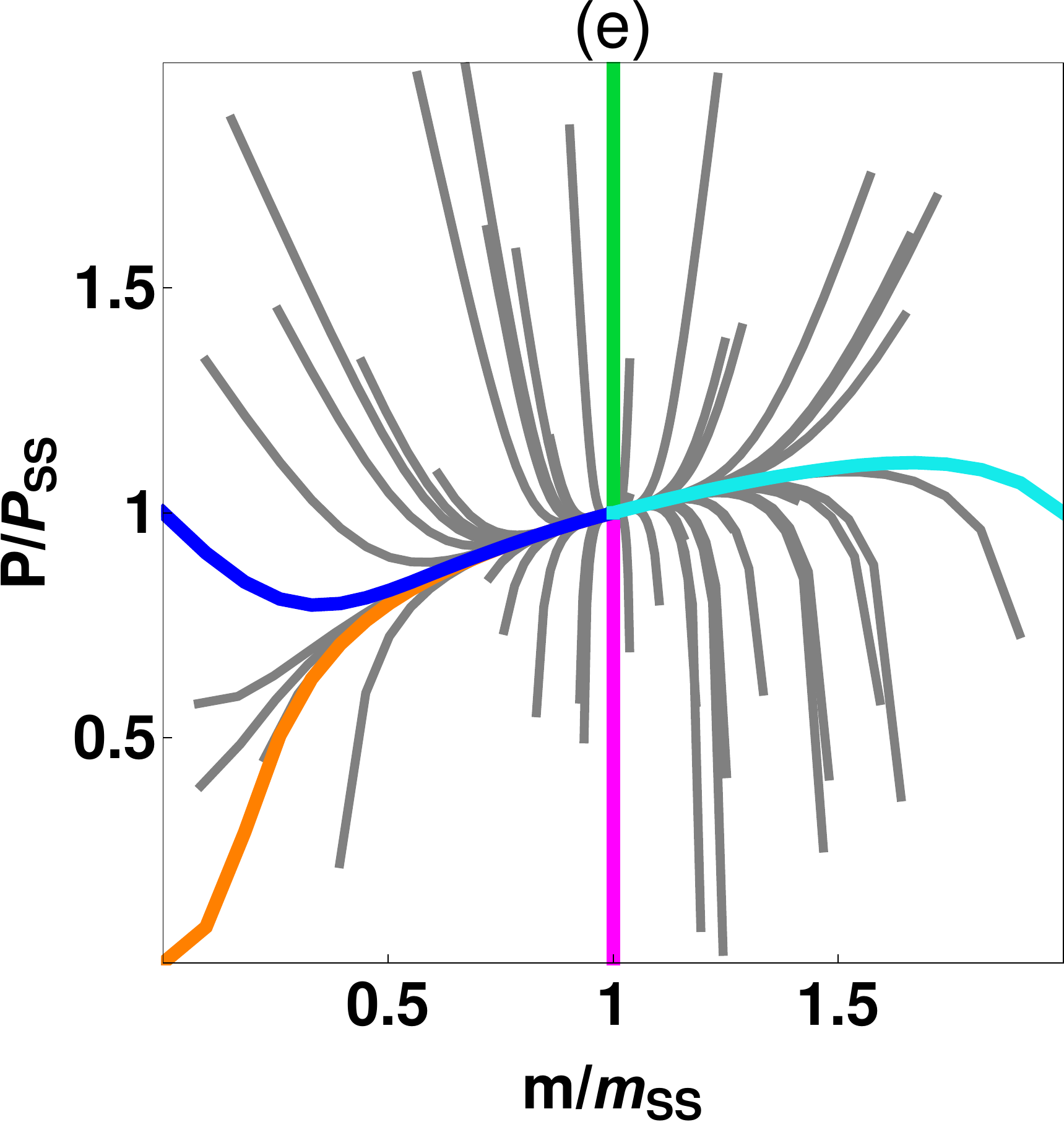}\quad
\includegraphics[width=2in]{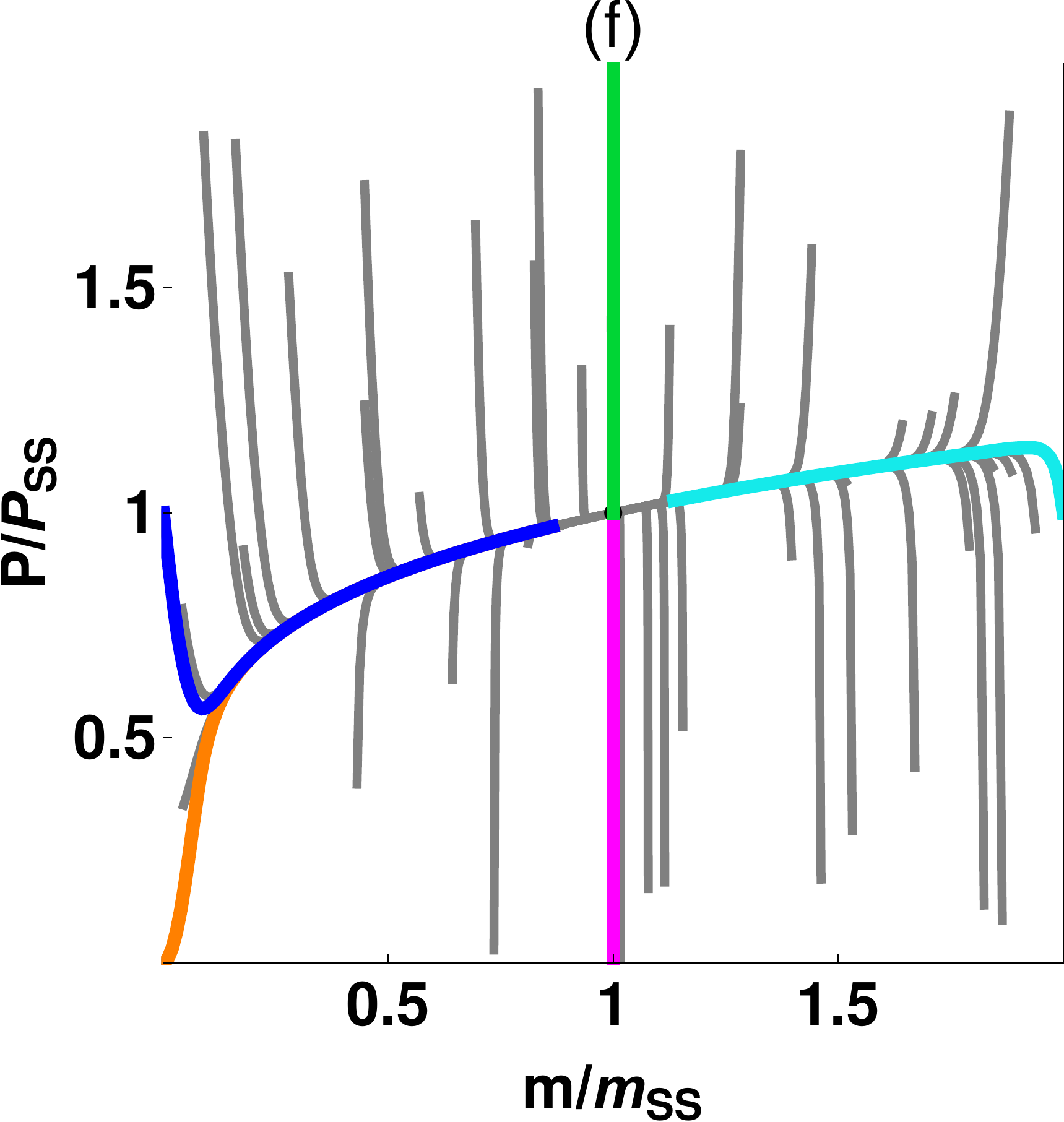} 
\medskip
\includegraphics[width=2in]{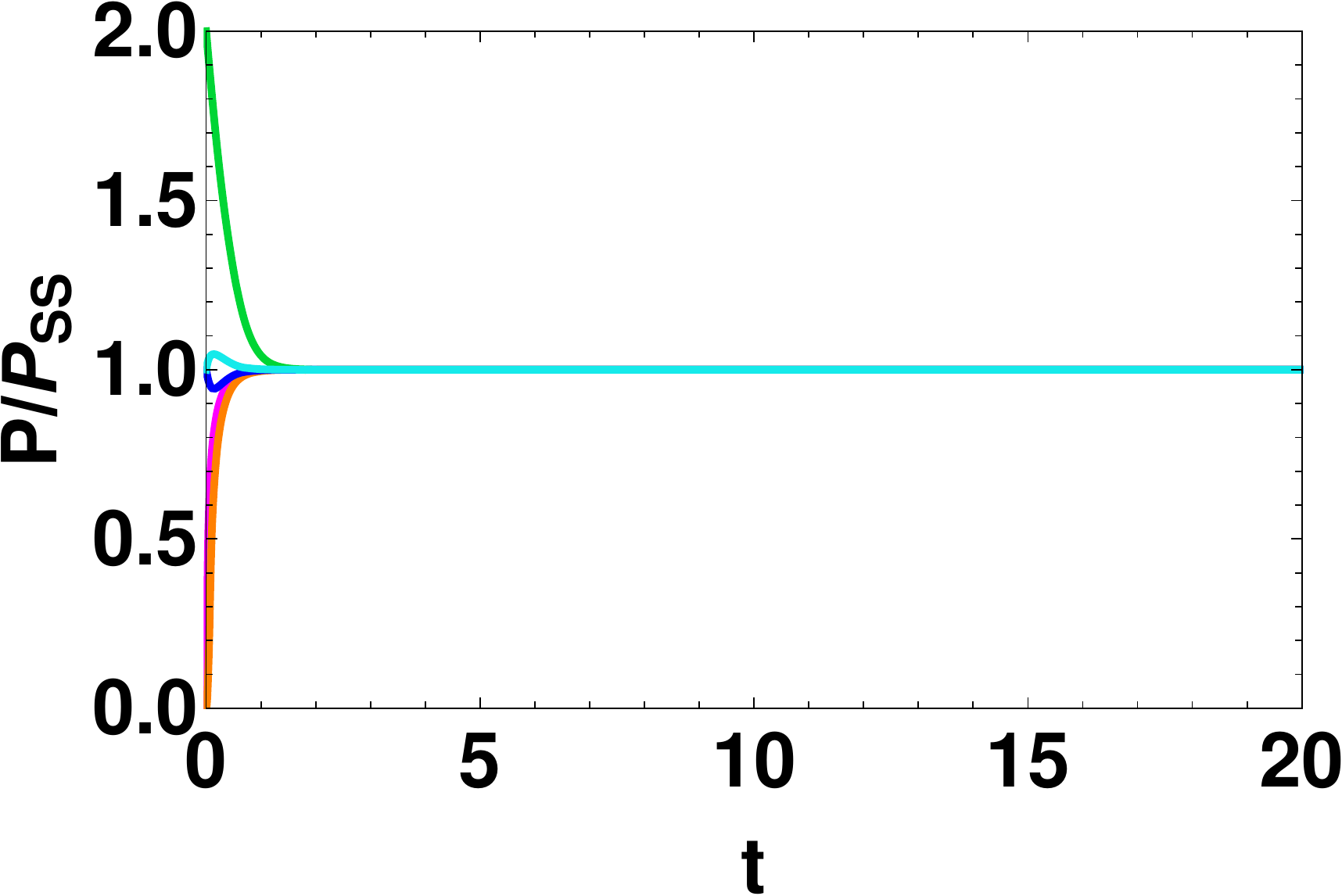}\quad
\includegraphics[width=2in]{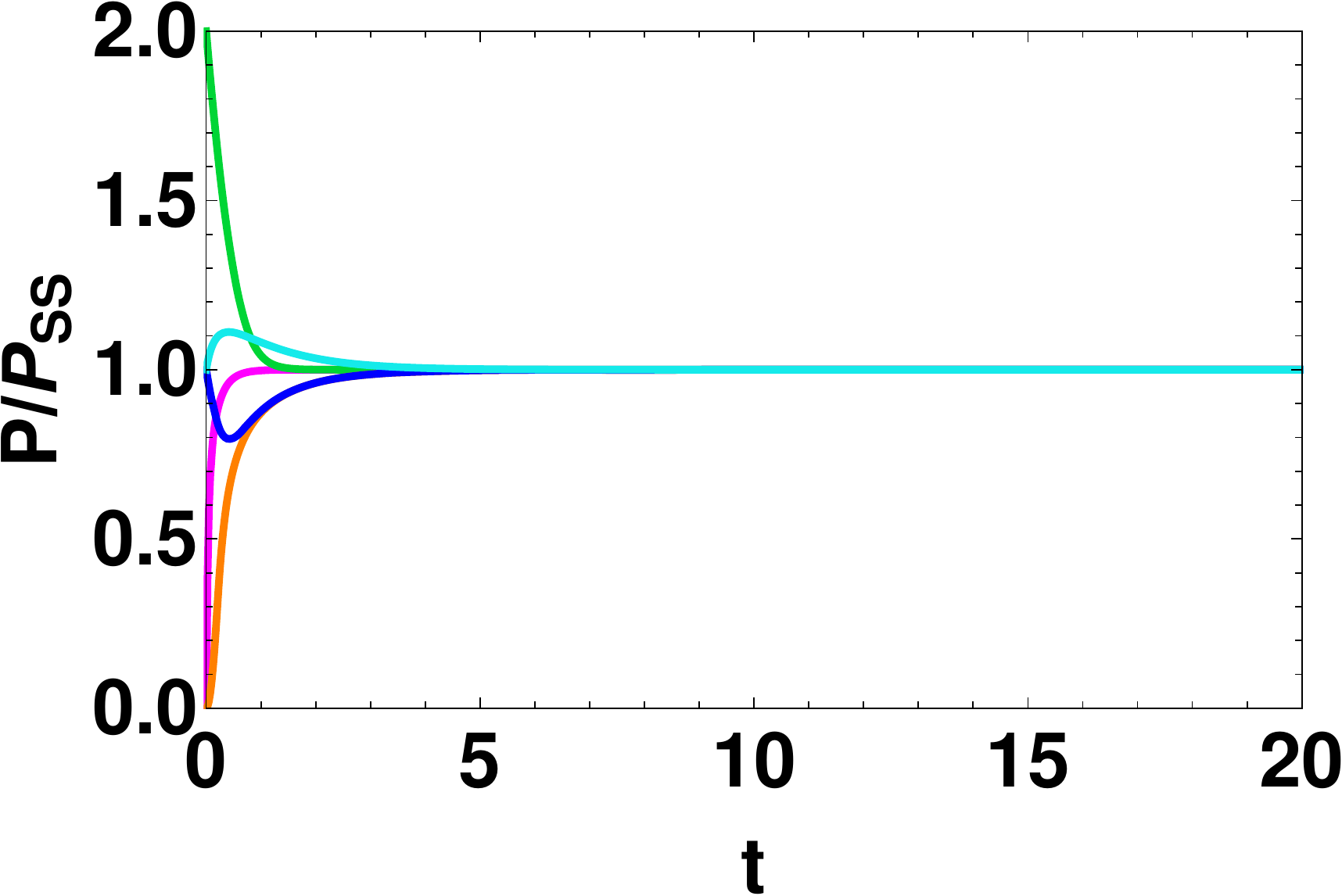}\quad
\includegraphics[width=2in]{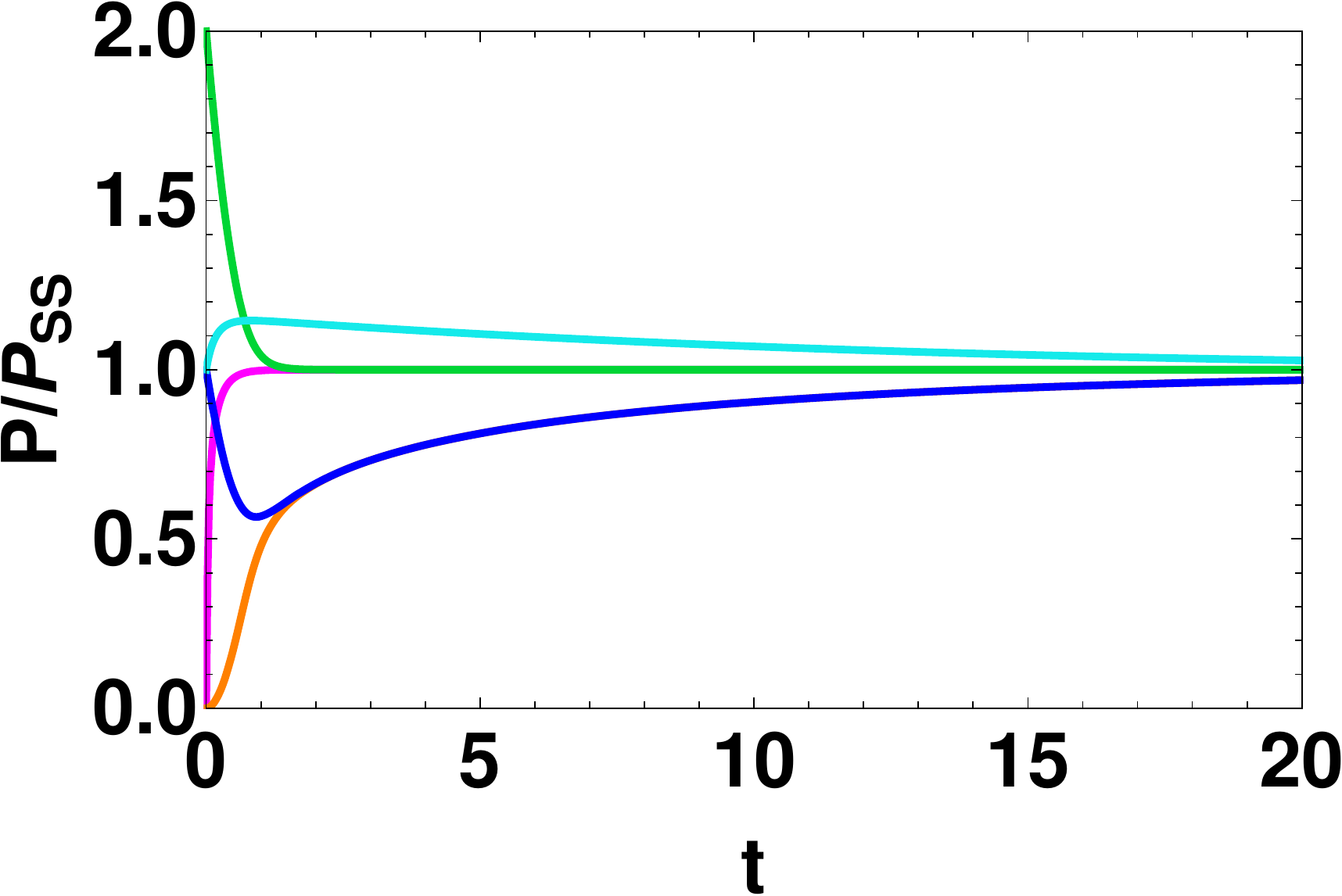} 
\end{center}
\end{figure}

\newpage
\begin{table}[H]
\caption{{Parameters used for Fig. \ref{tableIIfigures}. In
all cases, $k_2$ is fixed;  $k_1$ varies between linear and feedback systems by the factor $X_{SS}$. }}
\begin{center}
  \begin{tabular}{||c||c|c|c||}
    \hline
   \, & vertical, linear & focus, linear & diagonal, linear \\      \hline
   $k_1$ & $\frac{2}{17}$ & $\frac{2}{17}$ & $\frac{2}{17}$  \\
   $k_2$ & 1 & 1 & 1  \\
   $k_3$ & 3.4 & 34 & 340 \\
   $k_4$ & 0.1 & 1 & 10  \\
   $K$ & - & - & -  \\
   $N$ & - & - & -  \\
   $X_{SS}$ & 1 & 1 & 1 \\ \hline
   $m_{SS}$ & $\frac{2}{17}$ & $\frac{2}{17}$ & $\frac{2}{17}$ \\
   $P_{SS}$ & 4 & 4 & 4 \\
   $k_2/k_4$ & 10 & 1 & $\frac{1}{10}$ \\
   $\tau_1$,\space $\tau_2$, \space $\tau_3$  & 6.93, 6.93, 7.98 &
 0.693, 0.693, 1.68 & 0.0693, 0.0693, 0.798 \\ \hline \hline
   \, & vertical, feedback & focus, feedback & diagonal, feedback \\     \hline
   $k_1$ & 2 & 2 & 2 \\
   $k_2$ & 1 & 1 & 1 \\
   $k_3$ & 3.4 & 34 & 340 \\
   $k_4$ & 0.1 & 1 & 10 \\
   $K$ & 4 & 4 & 4 \\
   $N$ & 2 & 2 & 2 \\
   $X_{SS}$ & 17 & 17 & 17 \\ \hline
   $m_{SS}$ & 2 & 2 & 2 \\
   $P_{SS}$ & 4 & 4 & 4  \\
   $k_2/k_4$ & 10 & 1 & $\frac{1}{10}$ \\
   $\tau_1$,\space $\tau_2$, \space $\tau_3$  & 3.11, 0.361, 1.00 &
0.311, 0.0361, 0.297 & 0.0311, 0.00361, 0.106 \\ \hline \hline
   $\tau_{linear}/\tau_{feedback}$ & 2.23, 19.18, 7.97 & 2.23, 19.18, 5.65 & 2.23, 19.18, 7.56 \\ \hline \hline
  \end{tabular}
\end{center}          \label{TableII}
\end{table}

\newpage
\begin{figure}[H]
\caption{{Normalized phase portraits and timecourses for $P/P_{SS}$  for Table II.}\label{tableIIfigures}}
\begin{center}
\includegraphics[width=2in]{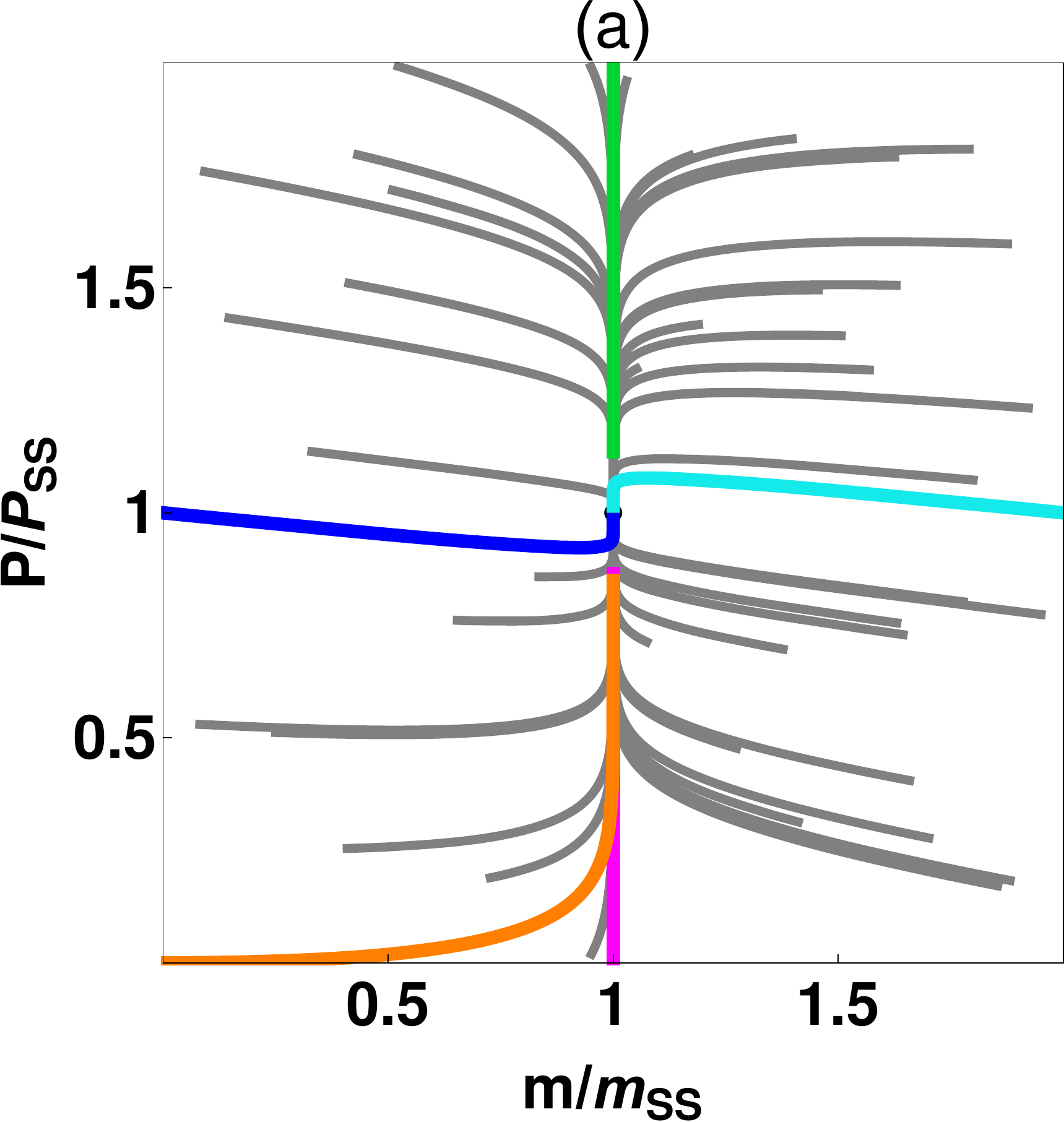}\quad
\includegraphics[width=2in]{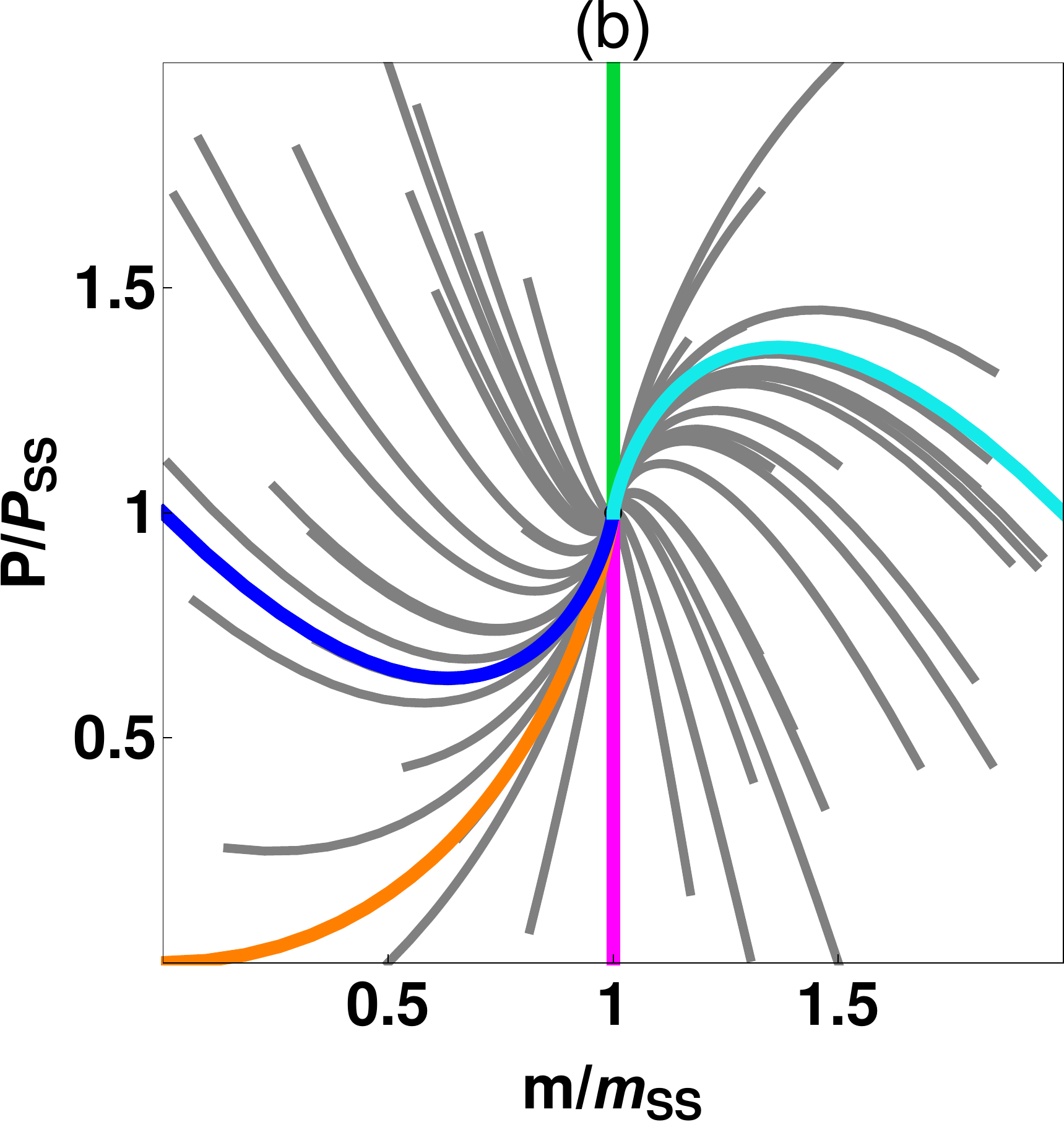}\quad
\includegraphics[width=2in]{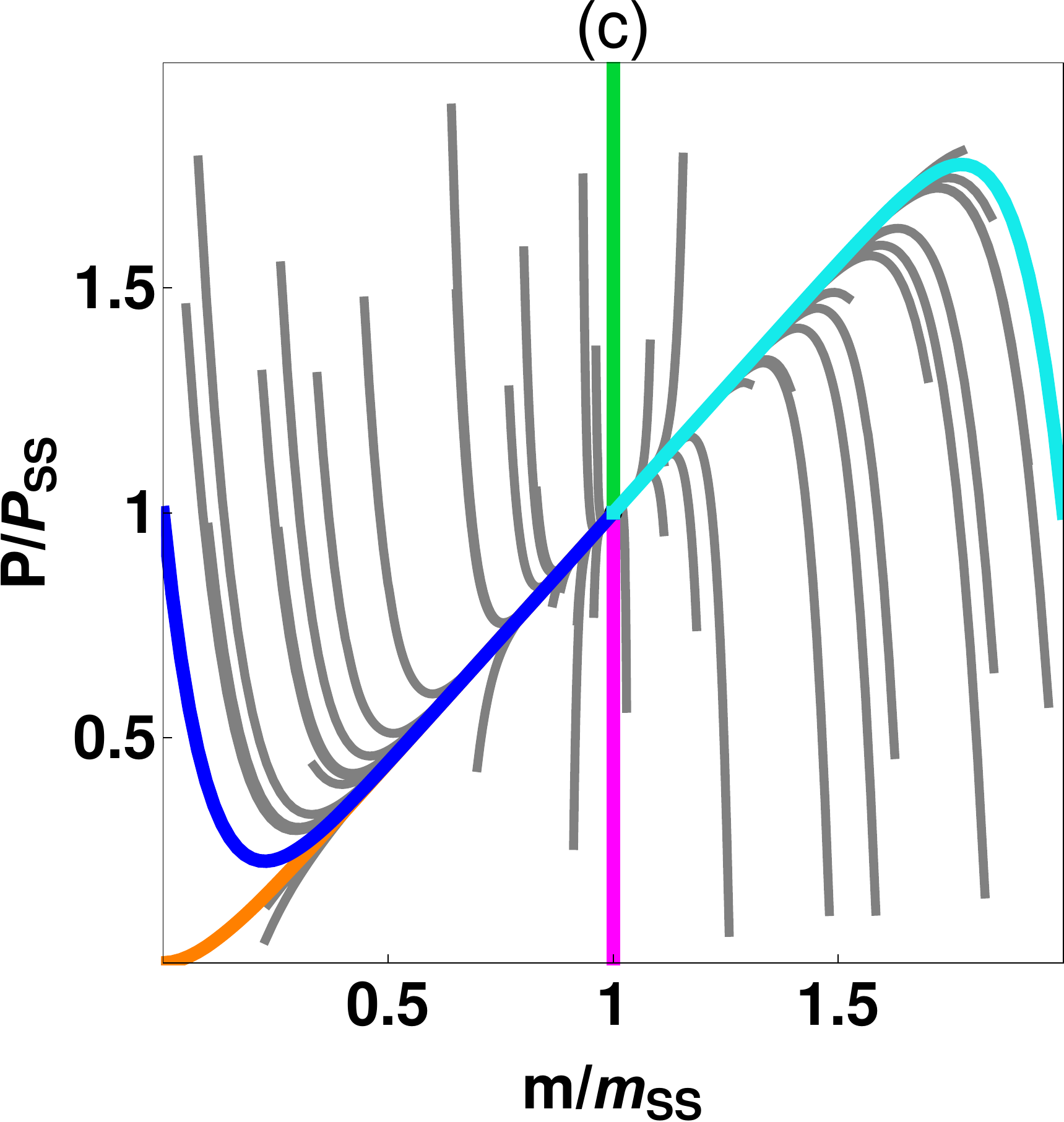} 
\medskip
\includegraphics[width=2in]{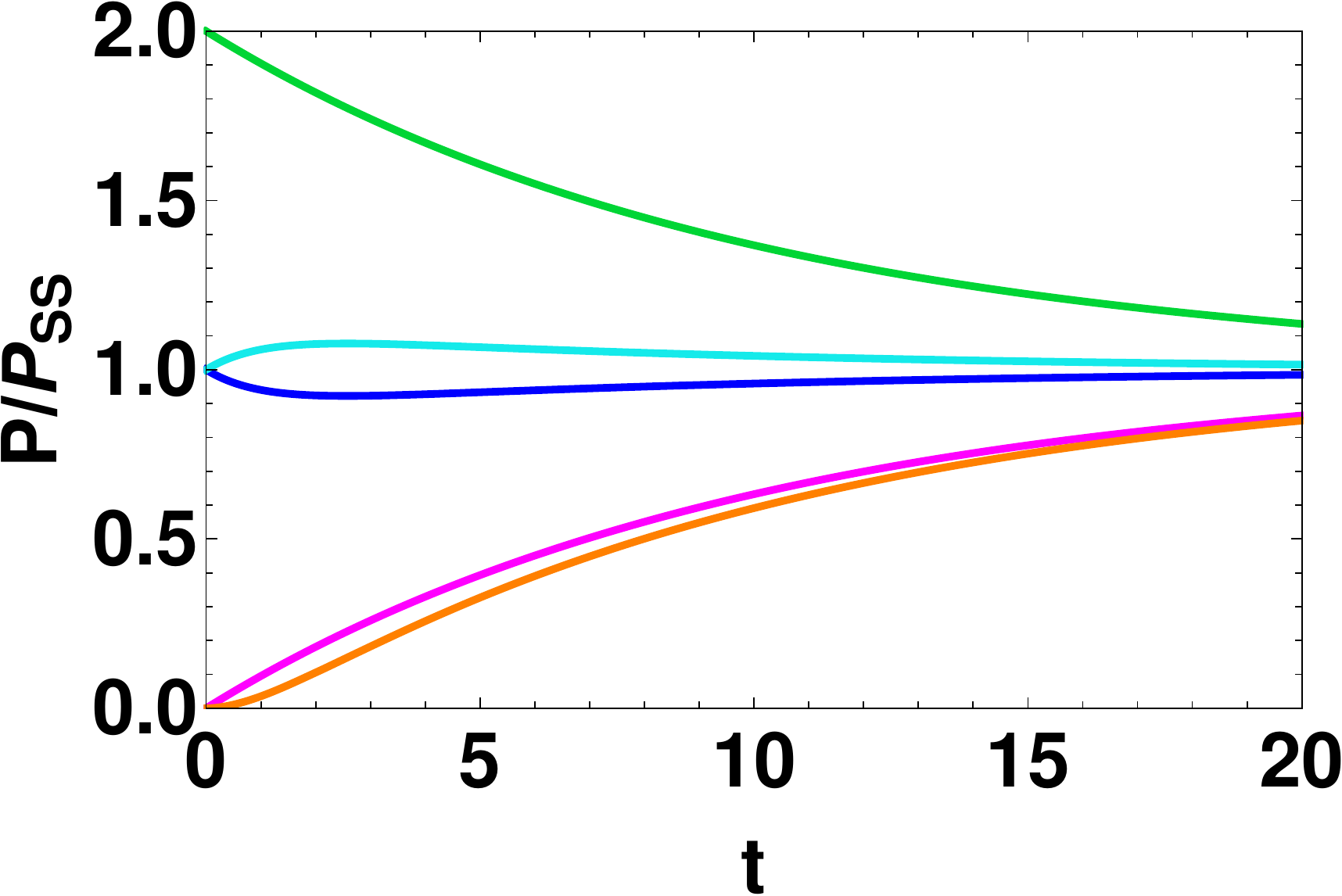}\quad
\includegraphics[width=2in]{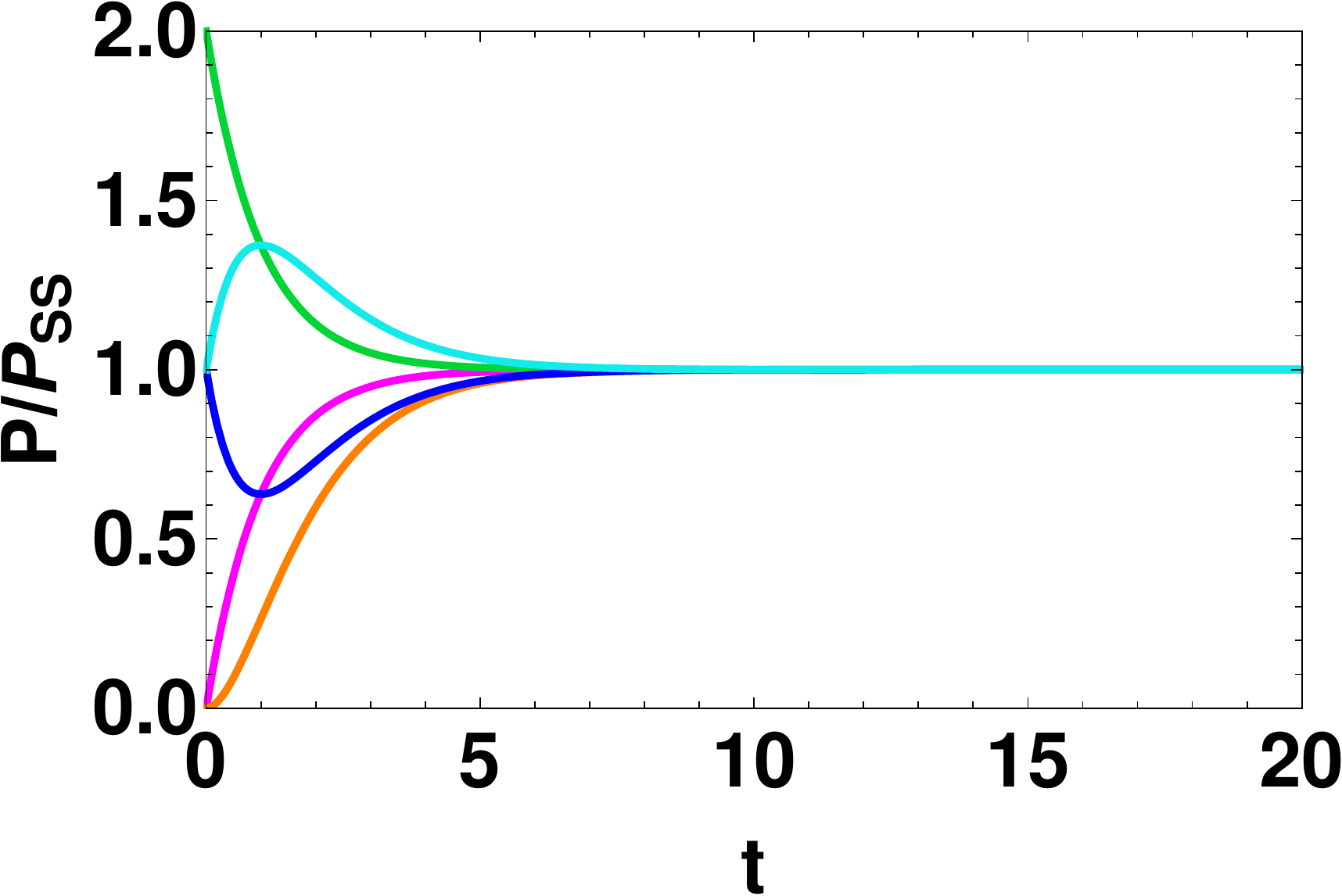}\quad
\includegraphics[width=2in]{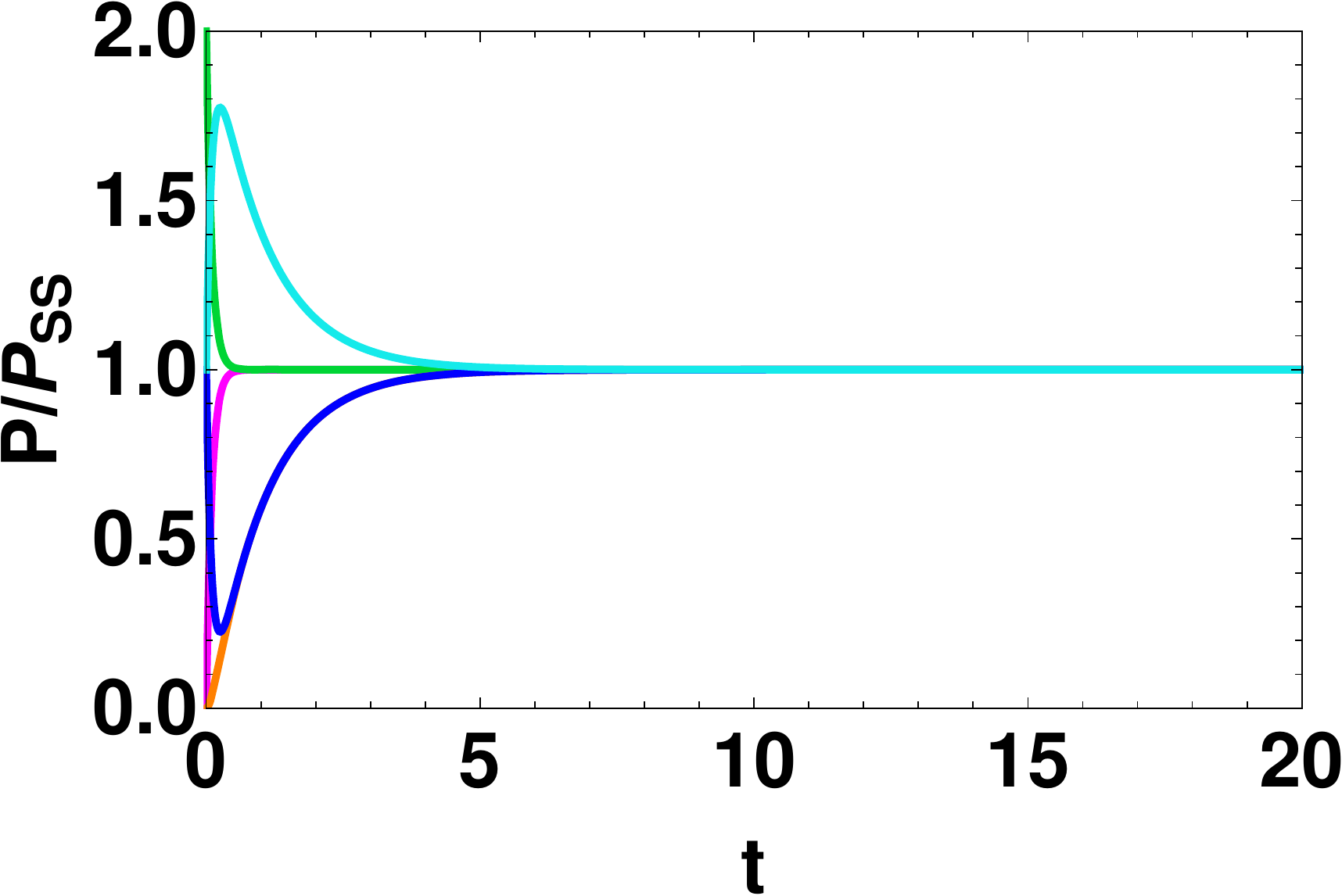} 
\medskip
\vspace{0.5in}
\includegraphics[width=2in]{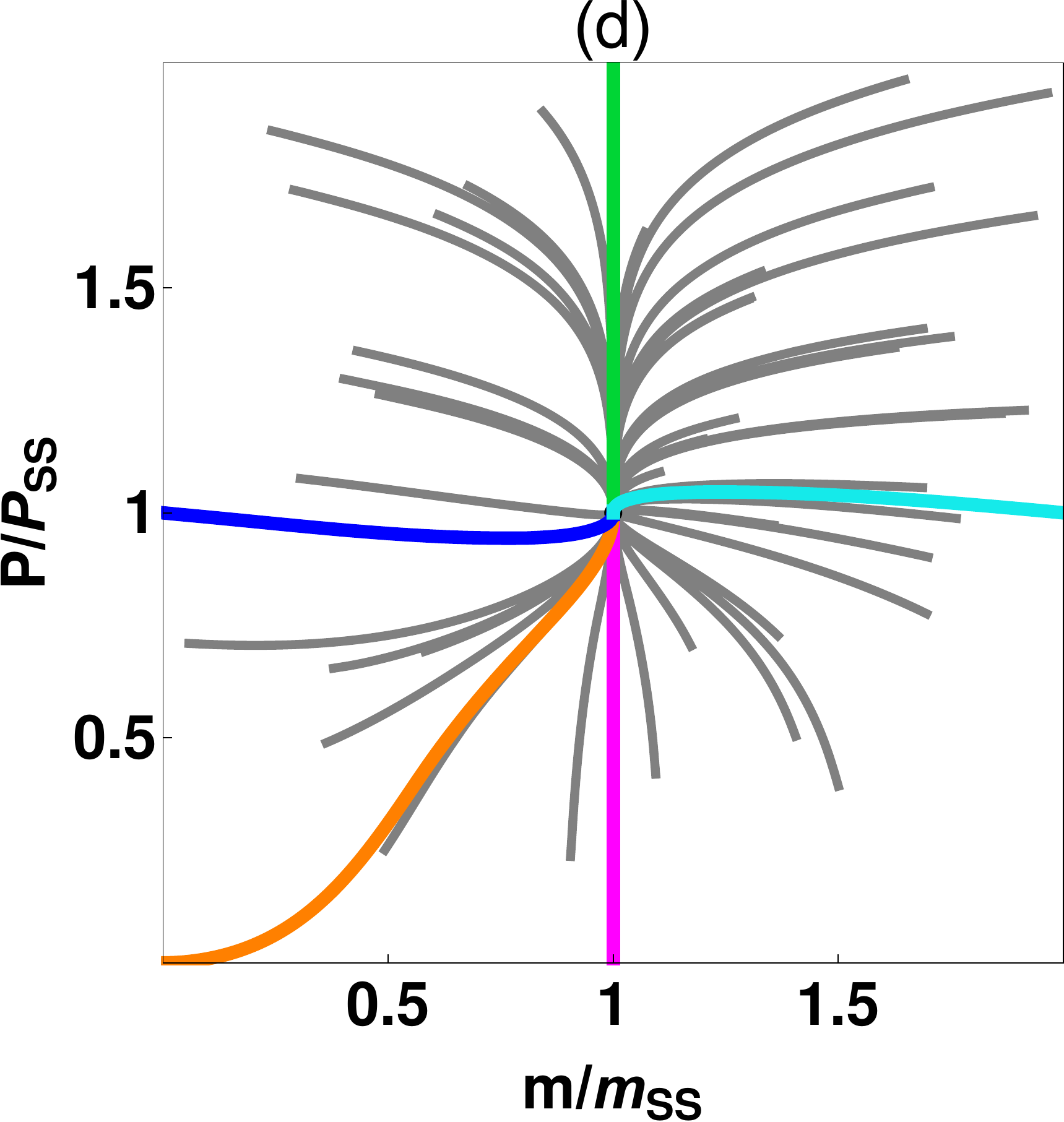}\quad
\includegraphics[width=2in]{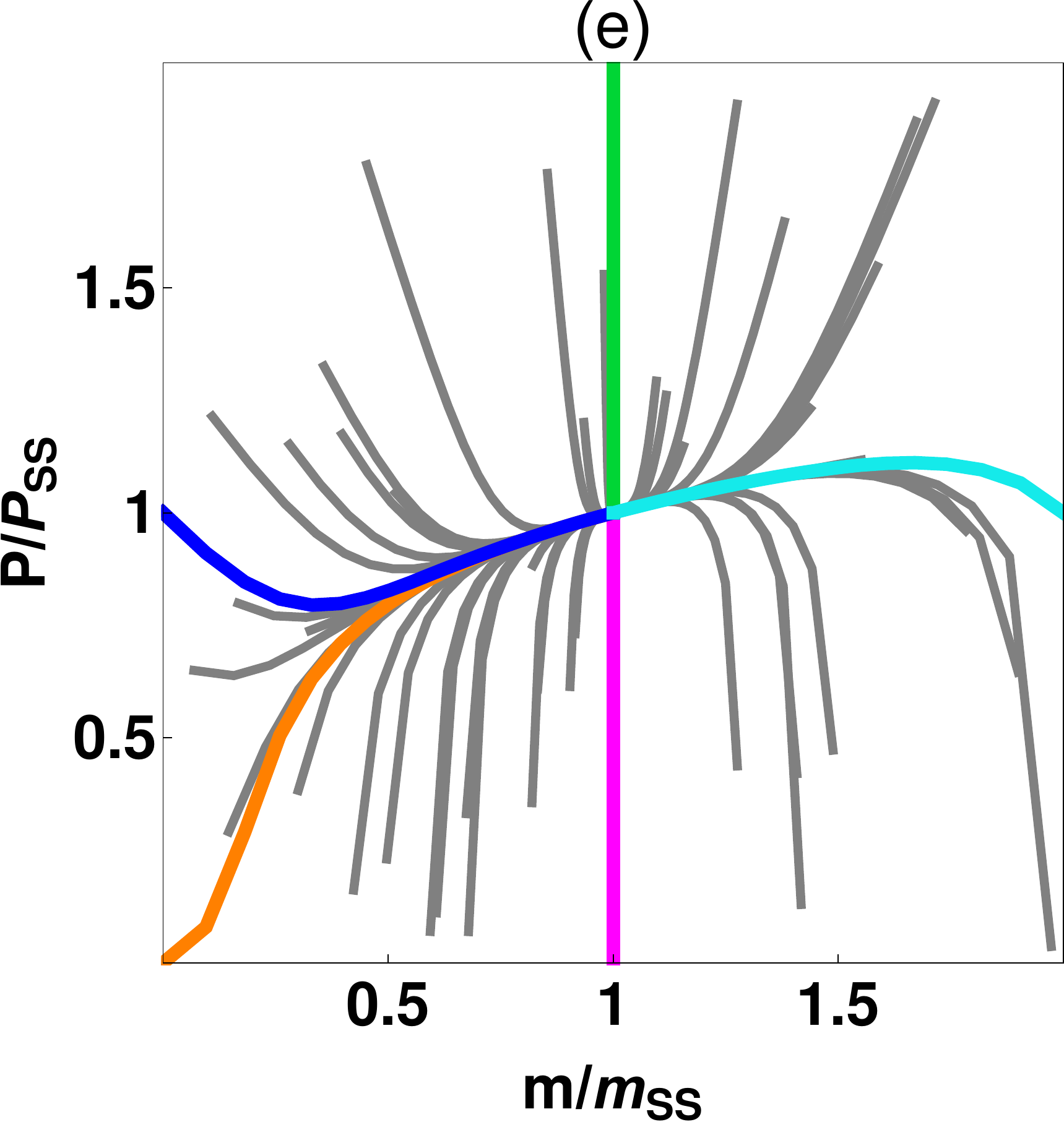}\quad
\includegraphics[width=2in]{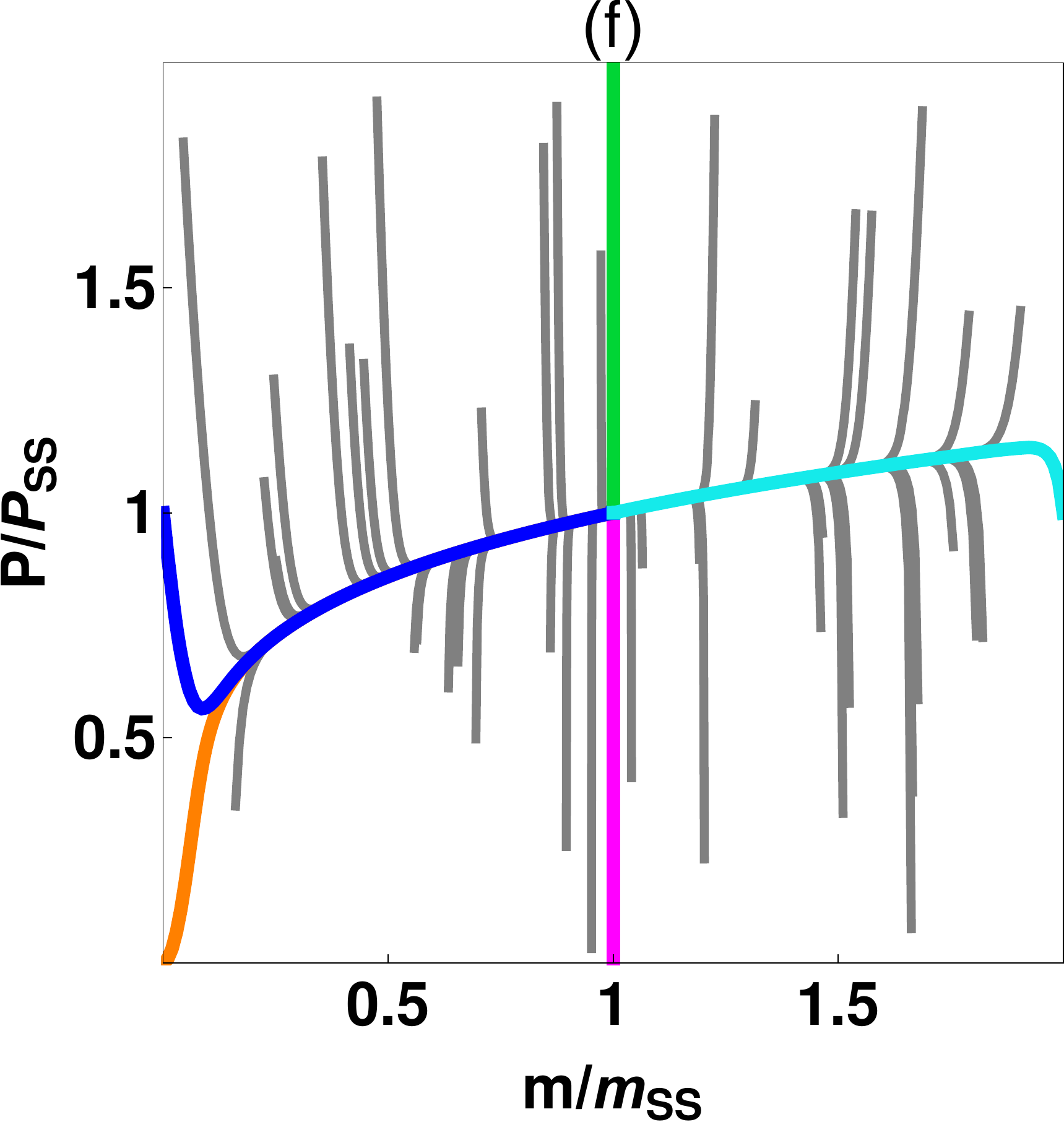} 
\medskip
\includegraphics[width=2in]{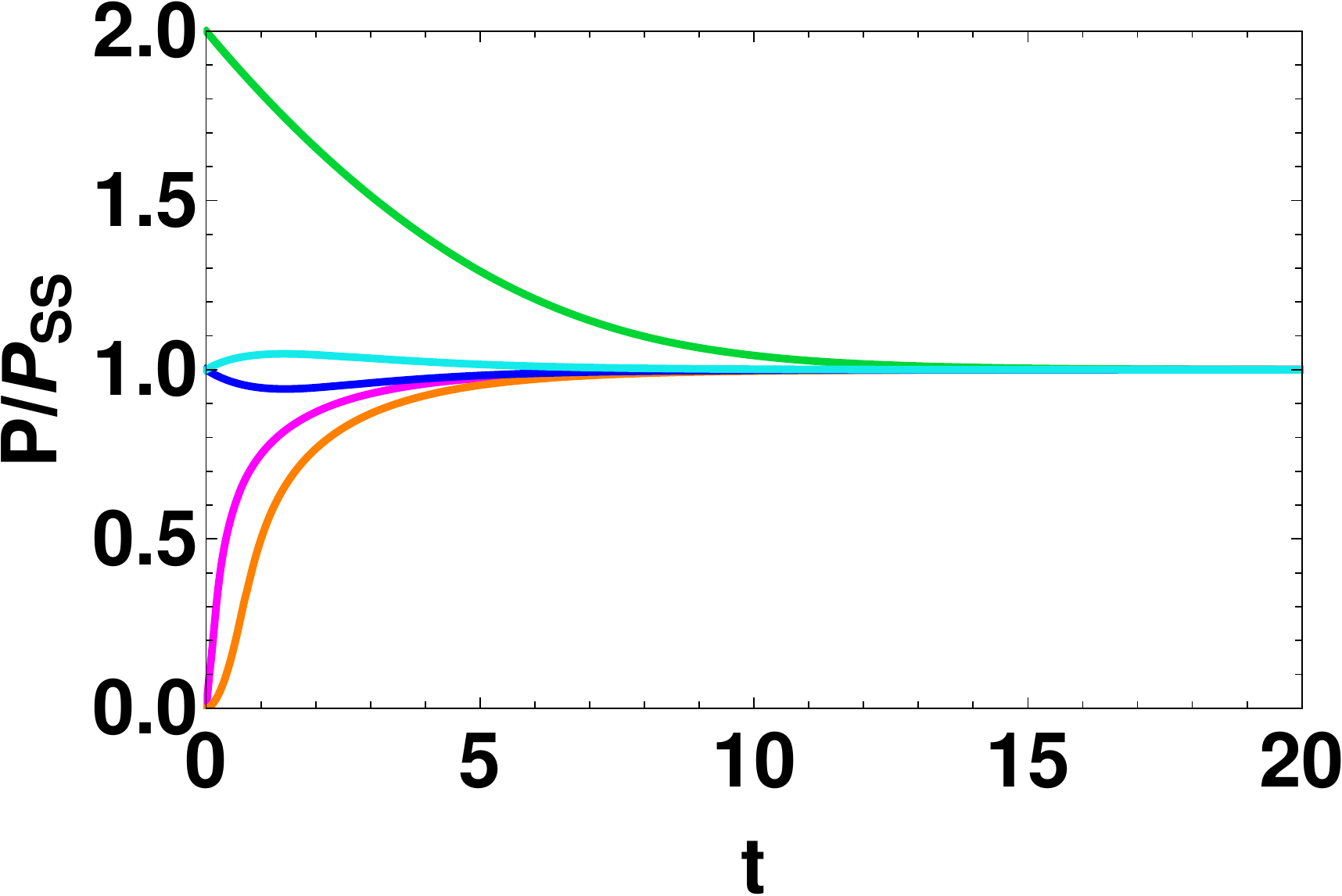}\quad
\includegraphics[width=2in]{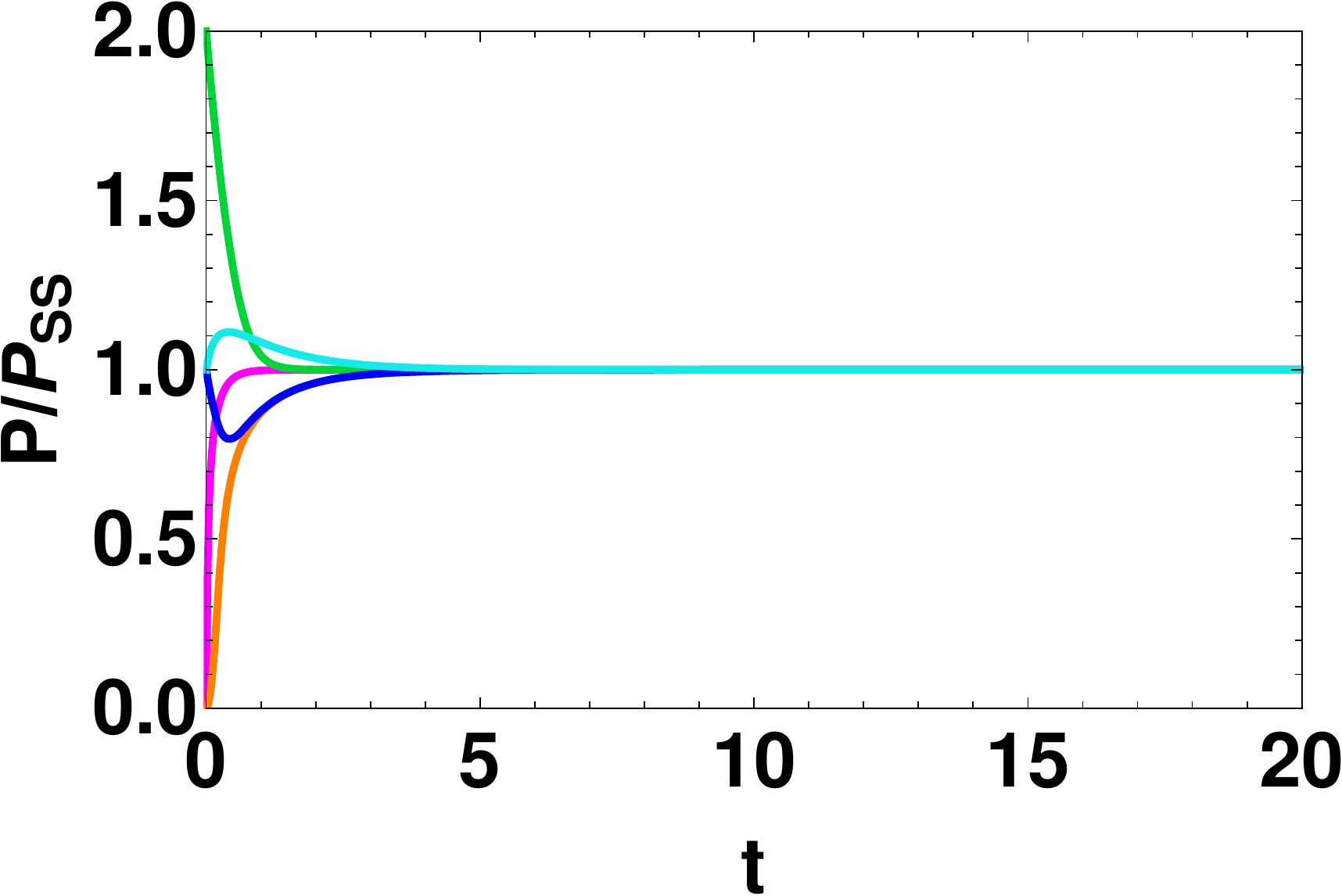}\quad
\includegraphics[width=2in]{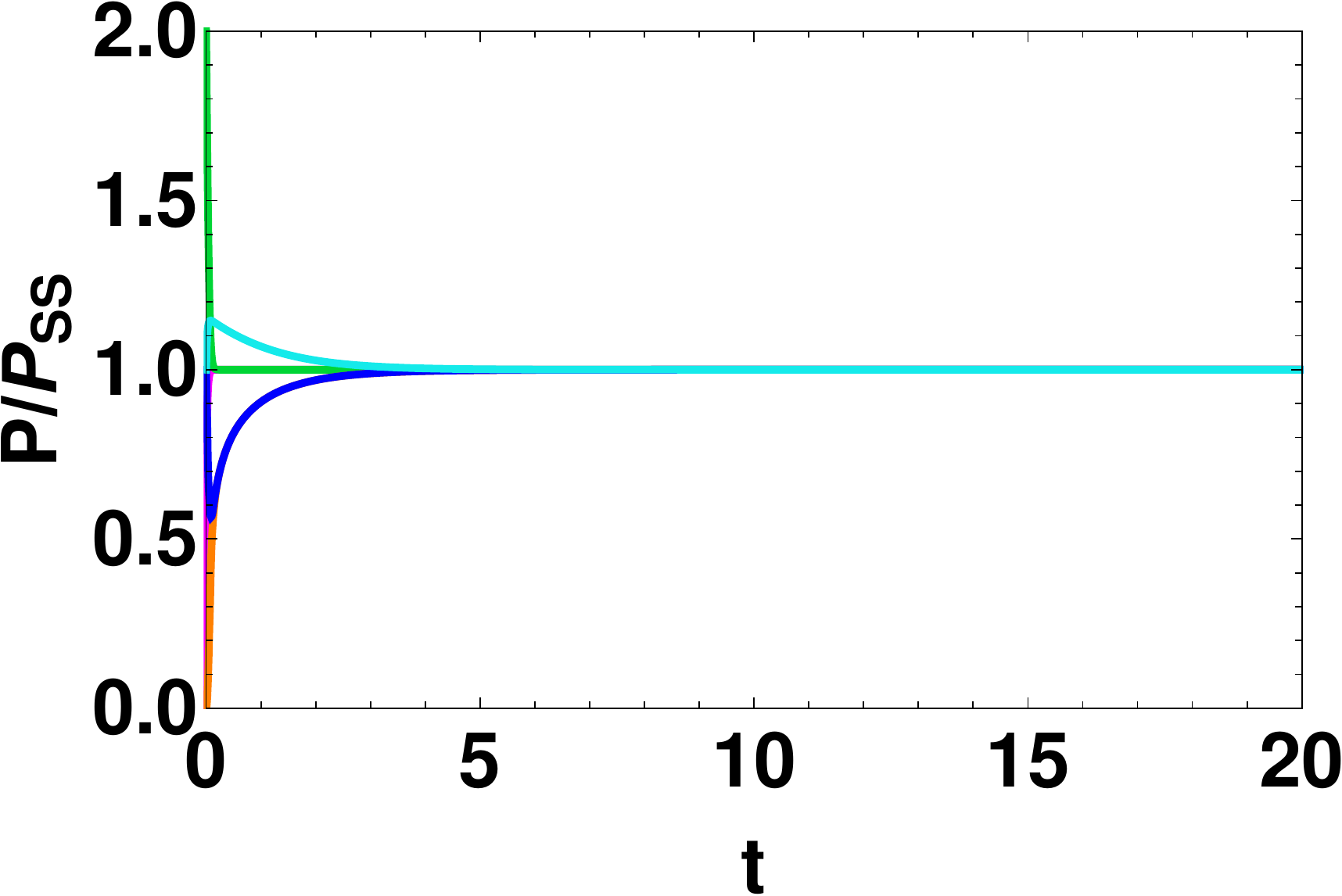} 
\end{center}
\end{figure}

\newpage
\,                                             

 The computational dynamics are shown in  Figs. \ref{tableIfigures}, \ref{tableIIfigures}, each figure showing the phase portraits for the linear systems with color-coded trajectories (top row), the time response of the $P$ concentration for various initial perturbations (second row), the phase portraits for the systems with feedback (third row), and the time response for the systems with feedback (fourth row).   It is immediately evident that the phase portraits in all cases depend crucially on the ratio of degradation rates  $k_2/k_4$.    It is also immediately clear that there are similarities -- in fact, exact invariances -- between corresponding phase portraits, linear and feedback, in Figs. \ref{tableIfigures}, \ref{tableIIfigures}.    By invariance we mean equivalence between the dynamical flow throughout the normalized portrait.  

The invariance of   linear phase portraits  between Figs. \ref{tableIfigures}, \ref{tableIIfigures} is easily explained by the fact that the time-dependent dynamics in the linear systems depends only on $k_2$ and $k_4$, as seen in Eq. \ref{lineartimedep}.    The explanation for the invariance evident in the feedback phase portraits between Figs.   \ref{tableIfigures}, \ref{tableIIfigures} is not so evident, since by 
from Abel's Impossibility Theorem \cite{abel}, we do not have exact analytic solutions, for either the concentrations $P, m$ or their derivatives in the rate equations, from which to make an invariance argument.  The surprise disappears when one examines the corresponding parameter sets in Tables I, II.  The invariances are between systems in which all the parameters $k_1 \cdots k_4$ are multiplied by a common factor of 10 or 1/10, i.e. in which there is merely an overall speedup or slowdown that obviously leaves the normalized phase portrait invariant.     This common scaling factor, obtained following the procedure of parameter variations outlined above, is due to the constraints we have applied on $P_{SS}$ and $m_{SS}$, together with  the relations   in Eqs. \ref{sssolution}, \ref{sssolutionfb}.

However, there is a further, genuine invariance puzzle that we will just touch upon here.    As noted already, the linear system is invariant under any parameter change that maintains $k_2/k_4$.   A surprising fact we have found is that the feedback system is  also  invariant under any parameter change that maintains both $k_2/k_4$ and $k_1/k_2 \times k_3/k_4$.  We will not discuss this much further here, as we do not  yet completely understand it.  We find that when we do vary $k_1/k_2 \times k_3/k_4$, the perturbation to the phase portrait is rather mild.   This ``symmetry" of the kinetic equations clearly merits future exploration.  In fact, these invariance and near-invariance properties are an important prediction of our model that could be tested in real biological systems with translational autoregulation.  

\subsection{Autoregulatory power: Time response of the linear and feedback systems}

Now we will examine the representative individual trajectories in Figs. \ref{tableIfigures}, \ref{tableIIfigures} with a view toward sizing up their powers of autoregulation.   The key criteria for regulation have to do with the response of the system under a perturbation -- of whatever cause -- foremost, the behavior of the protein concentration $P$; secondarily, the mRNA concentration $m$; with consideration of time and metabolic factors.      We will often make reference to the color-coding of the trajectories in the figures.

We focus first on the three phase space portraits of the linear system in Fig. \ref{tableIfigures}(a)-(c), associated with the parameters in Table \ref{TableI}; then on the three portraits of the feedback system in (d)-(f).  For each of the three  systems in each row, we have chosen trajectories corresponding to perturbed conditions that might be of prime importance, as touched upon in Section \ref{prototypes}. To reiterate from Section \ref{prototypes}:  the green trajectory in each figure  starts with  an excess of $P$ with $m$ fixed at the SS value.  The purple trajectory starts at reduced value of $P$ with $m$ fixed at  its  SS value.   The orange trajectory is for starting the system at zero values of both $m$ and $P$.  All of these trajectories eventually return to the SS.  Also shown are trajectories (cyan, blue)  in which the system is perturbed to deficient or excess values of $m$ with $P$ fixed at its  SS value.  The time response of the $P$ concentration for the various trajectories is shown under the phase portraits in the figures, in rows 2 and 4. 

The green and purple ``vertical trajectories" are each  identical among the three linear systems and also among the feedback systems (with  different response rates than the linear systems).   This is because 
the mRNA concentration is fixed at $m_{SS}$ along these trajectories, as follows mathematically from Eq. \ref{Pvertical}.      Moreover, the green and purple trajectories in each portrait have the same time dependence, again  from Eq. \ref{Pvertical}.  On the other hand, the other trajectories that have $m$ dependence differ greatly among the linear systems, and also among the feedback systems, showing the importance of the full two-species dynamics associated with the three classes and $k_2/k_4$.       It is evident from the time dependence of $P$ for the various trajectories, shown in the  second and fourth rows, that with feedback there is a great speedup in response times.   It is noteworthy that the purple and green trajectories differ in each feedback system portrait -- the green (excess $P$) is much slower than the purple (deficit of $P$).  We will examine the time response in more detail later shortly.

Fig. \ref{tableIIfigures}  shows the corresponding information for the systems in Table II where $k_4$ is varied, instead of $k_2$.  Not all of the statements regarding Fig. \ref{tableIfigures} apply in these cases.  Now the protein rate constants $k_3, k_4$ vary, and so do the time responses across rows 2 and 4 -- despite the exact invariance of the phase portraits between the two figures.  There is again a great speedup in the feedback systems.   Examination of the figures is probably more illuminating  than further verbal description.

We now turn to a quantitative measure of the response times and autoregulatory power of the various systems.  We take this, where it applies, to be the time $\tau_i$ for the $P$ concentration along a given trajectory to return halfway from its starting point to $P_{SS}$.  We give the values of  $\tau_1 \cdots \tau_3$ for the green, purple, and orange trajectories respectively at the bottom half of Tables I, II, together with the ratio of corresponding $\tau_{linear}/\tau_{feedback}$.  This ratio is an indication of response speedup due to the feedback; a larger ratio indicates a greater speedup.  A lot of regularities are evident among the $\tau_i$.  These are related to the symmetries,  invariances, and parameter ratios described above among the parameters and phase portraits, e.g. factors of 10.

We can see the great advantages to response time afforded by feedback as compared to the crude method in the linear system of just multiplying all the parameters by a common factor (thus providing an overall speedup -- at great metabolic cost).   Fig. \ref{tableIIfigures}a compared to Fig. \ref{tableIfigures}a shows the result of multiplying all the rates by a common factor.    Comparing Fig. \ref{tableIIfigures}a to Fig. \ref{tableIIfigures}d shows that a generally comparable speedup is obtained with feedback, by changing only one of the parameters $k_1$, the rate of transcription of the DNA,  instead of all of them.  This should be a great metabolic advantage, since the number of $m$ and especially of $P$ molecules produced by just one gene can be very large, ranging in the thousands \cite{Marcotte,mRNAEColi}.    Similar observations pertain to Fig. \ref{tableIfigures}c compared to Fig. \ref{tableIIfigures}c; and then Fig. \ref{tableIfigures}c compared to Fig.  \ref{tableIfigures}d.   The secret of the feedback regulation efficiency is basically that feedback, at relatively little cost, replaces degradation as the control mechanism, while sacrificing little in speed.  The ``brakes" are applied selectively, only as needed.

There is an interesting partial exception to these statements.  Note that the green trajectories (relieving excess $P$) in the feedback systems show a much slower response rate than the purple trajectories (relieving deficient $P$).  This might seem surprising in that the control scheme depicted in Fig. \ref{activatorrepressor} is based on jamming the production of $P$.   From the computed dynamics, apparently, greater inhibition of $P$ production at excess concentrations is more effective than lesser inhibition at deficient concentrations.

\section{Discussion and conclusions}

We end with a summing up, and a prospectus for future application of kinetic modeling of autoregulation  to important biological systems and problems, such as chemotherapy  of many cancers, briefly discussed below.  We have developed a two-component kinetic model for translational autoregulation, with mRNA concentration $m$ and protein concentration $P$ as dynamical variables.  The basic linear system gives a ``barebones" model with regulation of the system toward the target steady state.    However, the linear model is extravagantly wasteful:  to speed up control against perturbations away from the steady state, multiple rate constants must be increased, generally at great metabolic cost.  To attain more efficient control, feedback with cooperativity can be added to the model.  Adding feedback with cooperativity dramatically improves the autoregulatory response.   This attains the objective of control that is targeted and fast, yet efficient.     In general, in comparison to the linear systems, which depend entirely on degradation of $m$ and $P$ for control, the feedback achieves comparable speedup in response time at much lower metabolic cost.

For perturbation of $P$ alone, with $m$ maintained at its steady state value, the system is basically one-dimensional, with correspondingly simple kinetic relations.  However, with perturbation of both $m$ and $P$, the full two-dimensional description is essential, with a great variety of dynamical possibilities in the three basic phase space structures.   
Massive reorganization of the dynamics can occur when the system parameters are changed.   If the parameters become changed by the organism in response to a perturbation -- e.g. if either or both of the degradation rates $k_2, k_4$ are changed -- a new instance of the autoregulatory system is created which changes the entire behavior of the system.

To a large extent, the ratio of degradation rates $k_2/k_4$ for $m$ and $P$ governs the structure of the dynamical flow in the phase space portrait.   This means that linear systems are exactly invariant under all parameter changes that preserve $k_2/k_4$.  Systems with feedback are exactly invariant under parameter changes that preserve both $k_2/k_4$ and $k_1/k_2 \times k_3/k_4$ -- an interesting empirical ``symmetry" that is still under investigation.   Computation shows that systems with feedback change mildly under changes in  $k_1/k_2 \times k_3/k_4$   that preserve $k_2/k_4$.   
Consistent with the above statements, the autoregulatory response  for both the linear and the feedback systems depends greatly on the phase space structure associated with $k_2/k_4$.     These invariance properties are an important prediction of our model that possibly could be tested in real biological systems that have translational autoregulation.  

Both transcriptional and translational regulation are a widespread phenomenon of genetic control.  \cite{GlobalMammalian}.   In future work we plan to investigate interesting contrasts in the kinetics and dynamics of these types of autoregulation.  One of the pioneering examples of translational autoregulation involves the protein gp32 in replication of the virus T4 during infection of {\it E. coli} \cite{vonHippel1982}.  We hope to analyze a specific example of our model built with parameters derived as much as possible from experiment.  

This paradigmatic translational control mechanism has been found to be important in  cells \cite{Kozak,EukaryoticTransReg,Latchman}.    An  example  appears to be autoregulation of thymidylate synthase (TS), an enzyme which is very important in both normal and cancerous cells, the latter including many of the most common cancers.  The TS autoregulation is believed \cite{Chu1989TS,Chu2010} to be crucial in the development in cancer cells of resistance against chemotherapy drugs, many of which (e.g. 5-fluorouracil) specifically bind to TS in order to hinder DNA synthesis in cancer cells or induce apoptosis. This is very much like our purple trajectories in Figs. \ref{tableIfigures}, \ref{tableIIfigures}.    Then, it is possible that some reorganization takes place that would correspond to change  in one or more of the parameters of our model, associated with the unfortunate development of resistance to the drug.    We believe that building   quantitative kinetic schemes tailored to biological systems is essential for thorough understanding of the chemotherapeutic process, and perhaps even intervening in new ways.

We have tried to convey that it is pretty much hopeless to fully apprehend the dynamics of even so simple a model as ours here without the quantitative analysis of the full phase space dynamics.   This statement surely must carry over to more complex genetic regulatory networks -- underlining how daunting is a full understanding of their behavior.  This underlines the difficulty of realizing the promise of Waddington's heuristic notion of the ``epigenetic landscape"  \cite{Waddington,Ferrell,HuangKauffman}.
 
As a final remark, we make the observation that the autoregulatory model developed here does not function much like a ``program."  There is no logical scheme here, other than the tendency of the complete system to revert to the SS.  The ``intelligence" in the system is simply in the dynamics.    If biological systems function something like an operating system running programs, it happens at a larger or more complex scale of organization than the simple system here -- which would likely be a component of some such larger system.

\section{Appendix: Jacobian analysis}

Generally, for the differential equation set 

\begin{eqnarray}\frac{d x(t)}{d t} =
f_1(x, y), \frac{d y(t)}{d t} = f_2 (x, y)\end{eqnarray}

\noindent  a steady state ($x_0,y_0$) is defined as a point in the phase space where
$f_1=f_2=0$.  Near such a point, the dynamics can be approximated linearly as    \cite{Tabor} :
\begin{eqnarray}
\frac{d}{d t} \left( \begin{array}{c} x-x_0 \\ y-y_0 \end{array}
\right) = J \left( \begin{array}{c} x-x_0 \\ y-y_0
\end{array} \right) = \left( \begin{array}{cc} \frac{d f_1}{d x} &
\frac{d f_1}{d y} \\ \frac{d f_2}{d x} & \frac{d f_2}{d y} \end{array}
\right) \left( \begin{array}{c} x-x_0 \\ y-y_0
\end{array} \right)
\end{eqnarray}
where   $(x(t)-x_0, y(t)-y_0)$         is a vector and $J$  a matrix.   
\noindent Diagonalization of $J$ yields two eigenvalues $\lambda_1,
\lambda_2$ and their associated eigenvectors $\vec{V}_1, \vec{V}_2$.
At a stable steady state, $\lambda_1<0$ and $\lambda_2<0$, and the
solution of the linear system has the following form \cite{Tabor}:

\begin{eqnarray}
(x(t)-x_0, y(t)-y_0) = c_1 \vec{V}_1 e^{\lambda_1 t} + c_2 \vec{V}_2
e^{\lambda_2 t}
\end{eqnarray}
Hence, when $\vert \lambda_1 \vert$ $\gg$ $\vert \lambda_2 \vert$,
there is a natural separation of time scale, with the fast and slow
components along the directions of $\vec{V}_1$ and $\vec{V}_2$,
respectively. Note that the eigenvectors may not be orthogonal to each
other when $J$ is not symmetric, which is the case here.
In the no-feedback model, from the kinetic equations (\ref{lineareqns}) we obtain 

\begin{eqnarray}
\lambda_1 &&=-k_2, \quad \quad \lambda_2=-k_4 \\
V_1 &&= \{\frac{k_4 -k_2}{k_3}, 1\}, \quad \quad V_2 = \{0, 1\}
\end{eqnarray}

\noindent Note that in this case $V_2$ always points vertically, since
${d m}/{d t}$ does not depend on $P$ (while ${d P}/{d t}$
depends on $m$).   This has the consequence that trajectories on  left and right of the vertical are
strictly separated, as is apparent from the phase space portraits.   By considering different values of $k_1-k_4$, the
dynamics can be divided into the 3 categories depending on the ratio
of $k_2/k_4$.

\begin{acknowledgement}
We would like to thank Pete von Hippel and members of his group seminar for many stimulating discussions about translational autoregulation, especially of gp32 in the T4 - E. coli  system.  
\end{acknowledgement}

\providecommand{\latin}[1]{#1}
\makeatletter
\providecommand{\doi}
  {\begingroup\let\do\@makeother\dospecials
  \catcode`\{=1 \catcode`\}=2 \doi@aux}
\providecommand{\doi@aux}[1]{\endgroup\texttt{#1}}
\makeatother
\providecommand*\mcitethebibliography{\thebibliography}
\csname @ifundefined\endcsname{endmcitethebibliography}
  {\let\endmcitethebibliography\endthebibliography}{}

\newpage
\begin{center}
\vspace{2in}\hspace{-1in}
\includegraphics[width=5in]{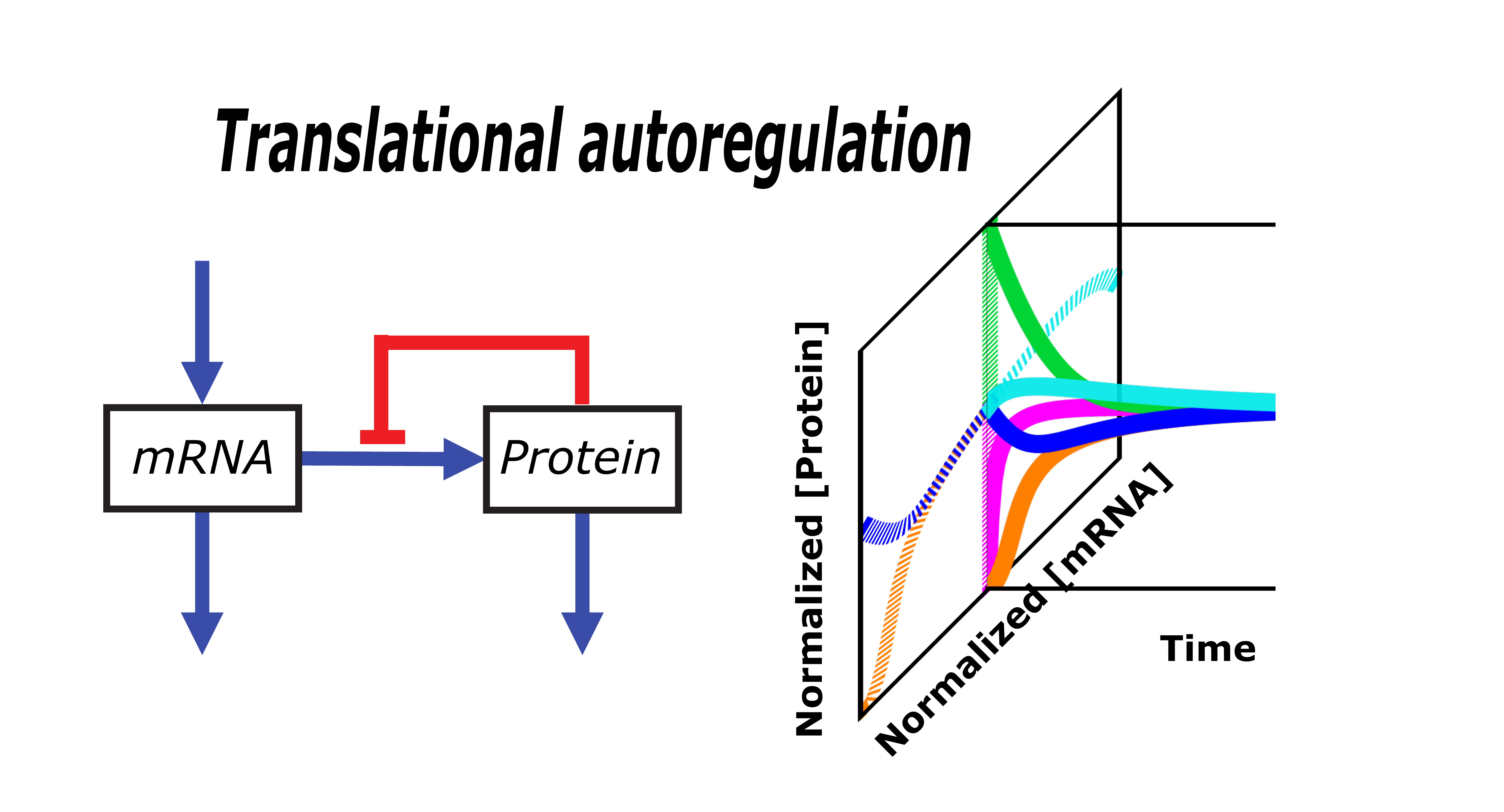} 
\end{center}

\end{document}